\documentclass[11pt]{article}

\usepackage[margin=1in]{geometry}
\usepackage{setspace}
\usepackage{amsmath,amssymb,amsfonts,bm}
\usepackage{graphicx}
\usepackage{caption}
\usepackage{subcaption}  
\usepackage{booktabs,multirow,array}
\usepackage{float}
\usepackage{rotating}
\usepackage{tikz}
\usetikzlibrary{decorations.pathreplacing}
\usepackage{adjustbox}
\usepackage{algorithm}
\usepackage{algpseudocode}
\usepackage{enumitem}
\PassOptionsToPackage{table}{xcolor}
\usepackage{xcolor}      
\usepackage{colortbl} 
\definecolor{Gray}{gray}{0.9}
\usepackage{epstopdf} 
\usepackage{orcidlink}
\usepackage{hyperref}
\usepackage[numbers,sort&compress]{natbib} 

\newtheorem{proposition}{Proposition}
\newtheorem{theorem}{Theorem}
\DeclareMathOperator*{\argmax}{arg\,max}

\global\long\def\mbA{\mathbf{A}}%
\global\long\def\mbB{\mathbf{B}}%
\global\long\def\mbC{\mathbf{C}}%
\global\long\def\mbD{\mathbf{D}}%
\global\long\def\mbI{\mathbf{I}}%
\global\long\def\mbO{\mathbf{O}}%
\global\long\def\mbP{\mathbf{P}}%
\global\long\def\mbU{\mathbf{U}}%
\global\long\def\mbV{\mathbf{V}}%
\global\long\def\mbx{\mathbf{x}}%
\global\long\def\mbX{\mathbf{X}}%
\global\long\def\mby{\mathbf{y}}%
\global\long\def\mbY{\mathbf{Y}}%
\global\long\def\mbZ{\mathbf{Z}}%
\global\long\def\bolGamma{\boldsymbol{\Gamma}}%
\global\long\def\bolSigma{\boldsymbol{\Sigma}}%
\global\long\def\bolPhi{\boldsymbol{\Phi}}%
\global\long\def\calI{\mathcal{I}}%
\global\long\def\calN{\mathcal{N}}%
\global\long\def\Cov{\mathrm{cov}}%
\global\long\def\cov{\mathrm{cov}}%
\global\long\def\Corr{\mathrm{corr}}%
\usepackage{algpseudocode}
\algnewcommand{\Inputs}[1]{
  \State \textbf{Inputs:}\vspace{2mm}
  \Statex \hspace*{\algorithmicindent}\parbox[t]{.8\linewidth}{\raggedright #1}
}
\algnewcommand{\Initialize}[1]{
  \State \textbf{Initialize:}\vspace{2mm}
  \Statex \hspace*{\algorithmicindent}\parbox[t]{0.8\linewidth}{\raggedright #1}
}
\algnewcommand{\Update}[1]{
  \State \textbf{Update:}\vspace{2mm}
  \Statex \hspace*{\algorithmicindent}\parbox[t]{.8\linewidth}{\raggedright #1}
}
\algnewcommand{\Output}[1]{
  \State \textbf{Output:}
  \Statex \hspace*{\algorithmicindent}\parbox[t]{.9\linewidth}{\raggedright #1}
}

\title{Dimension Reduction for Characterizing Sexual Dimorphism in Biomechanics of the Temporomandibular Joint}

\author{
  Sung Hee Park\orcidlink{0000-0002-4754-2611}\thanks{Email: \texttt{sunghee@wustl.edu}}$^{\enspace 1,2}$ \and
  Xin Zhang$^{1}$ \and
  Elizabeth Slate$^{1}$ \and
  Shuchun Sun$^{3}$ \and
  Hai Yao$^{3}$
}
\date{%
$^{1}$Department of Statistics, Florida State University, Tallahassee, FL, U.S.A.\\
$^{2}$Institute for Informatics, Data Science and Biostatistics, Washington University in St.\ Louis, MO, U.S.A.\\
$^{3}$Clemson\textendash MUSC Bioengineering Program, Department of Bioengineering, Clemson University, SC, U.S.A.\\[6pt]
September 2025
}

\begin{document}
\maketitle

\begin{abstract}
Sexual dimorphism is a critical factor in many biological and medical research fields. In biomechanics and bioengineering, understanding sex differences is crucial for studying musculoskeletal conditions such as temporomandibular disorder (TMD). This paper focuses on the association between the craniofacial skeletal morphology and temporomandibular joint (TMJ) related masticatory muscle attachments to discern sex differences. Data were collected from 10 male and 11 female cadaver heads to investigate sex-specific relationships between the skull and muscles. We propose a conditional cross-covariance reduction (CCR) model, designed to examine the dynamic association between two sets of random variables conditioned on a third binary variable (e.g., sex), highlighting the most distinctive sex-related relationships between skull and muscle attachments in the human cadaver data. Under the CCR model, we employ a sparse singular value decomposition algorithm and introduce a sequential permutation for selecting sparsity (SPSS) method to select important variables and to determine the optimal number of selected variables.
\end{abstract}


\section{INTRODUCTION}\label{M-sec:introduction}
Sexual dimorphism significantly influences human skull morphology and biomechanics, shaping our understanding of conditions like temporomandibular disorder (TMD). TMD affects $5-12\%$ of Americans, with an estimated annual cost of approximately \$4 billion \citep{stowell2007cost}. The relationship between temporomandibular joint (TMJ) muscle attachments and skull features is central to TMJ function and TMD development. TMD is multifactorial, with the morphology of the masticatory system contributing significantly to its development. Clinical studies have shown that women are more likely than men to develop TMD, with reported prevalence ratios ranging from 3:1 to 8:1 \citep{slade2011study}. Existing results in TMJ research are often reduced to summary statistics or rely on one-variable-at-a-time approaches \citep{coombs2019temporomandibular,she2021sexual}, which lack statistical efficiency and may obscure important associations. By integrating multimodal data from skull and muscle measurements—more accurately, craniofacial skeletal morphology and masticatory muscle attachment measurements—this work aims to provide more precise insights into TMJ biomechanics and craniofacial disorders.

\subsection{Motivating Data}\label{M-sec:data}
Motivated by a recent study on TMJ muscle attachment morphometry and musculoskeletal characterization \citep{she2018three}, we propose an integrative analysis framework that leverages the multivariate structure of multimodal craniofacial data. This approach is designed to identify associations that are highly sensitive to subject-level heterogeneity, explicitly accounting for sex differences. In the study by \cite{she2018three}, human cadavers and a custom surgical probe were used to quantify three-dimensional muscle attachment morphology. These data were then combined with cone beam computed tomography (CBCT) scans to explore their relationship with musculoskeletal modeling of the TMJ. The resulting dataset represents a valuable resource, as complete muscle shapes and orientations are essential for accurately characterizing TMJ biomechanics—yet such information cannot be directly obtained through current imaging technologies. The proposed integrative analysis aims to elucidate how different data modalities, such as muscle and skeletal measurements, interact and associate when conditioned on sex. 

Data were obtained from 21 cadaver heads (11 females, $73.6 \pm 12.8$ years; 10 males, $75.8 \pm 8.3$ years) without craniofacial abnormalities or TMD, as described in \cite{she2018three}. CBCT (voxel size $0.2 \times 0.2 \times 0.2 \enspace mm^{3}$) was used to reconstruct 3D craniofacial models, and dissections identified muscle attachment sites. After scanning with CBCT, solid 3D models of each head were reconstructed. Craniofacial anthropometric dimensions were measured from reconstructed 3D solid models of cadaver heads. TMJ muscle attachment morphometry was quantified using a co-registered CBCT and 3D digitization method \citep{she2021sexual}. A bounding-box approach defined attachment size (length, width, thickness, area), centroid coordinates, and orientation relative to anatomical planes. 
Measurements were made on eight TMJ muscle attachments. We focus on the temporalis origin (TO), which is critical for load-bearing tasks such as biting and chewing \citep{hylander1985temporalis}. 
The variables and dimensions of the TO and skull measurements are summarized in Table \ref{S-table:variables} of the Supplementary Materials.

\subsection{Statistical Problem Formulation and Related Work}
In our TMJ data analysis, let $\mbX \in \mathbb{R}^{p_{1}}$ denote the skull characteristic measurements extracted from the CBCT scans and $\mbY \in \mathbb{R}^{p_{2}}$ be the muscle attachment measurements. Then the problem can be statistically formulated as studying the relationship between $\mbX$ and $\mbY$ conditional on sex variable $Z\in\{1,2\}$. Despite rich statistical literature studying the associations between two sets of multivariate variables, notably the canonical correlation analysis (CCA) and its variants \citep{li2018general,mai2019iterative,shu2020d,wang2015inferring,witten2009penalized}, modeling their interrelationship conditional on a third set of variables is an important new research frontier in modern multivariate analysis. Here, statistical models and methods are needed for finding features of $\mbX$ and $\mbY$ whose associations differ by sex. Existing CCA-based approaches, in particular, aim to estimate a common pattern across sex groups. The proposed analysis framework therefore addresses this key limitation of CCA-based approaches and enables the discovery of novel, new insights from complex craniofacial  data.
 
In the multivariate analysis literature, dynamic association defined by conditioning on a third variable has been previously proposed. Most relatedly, liquid association is a concept originally proposed by \citet{li2002genome} to capture dynamic co-expression in two gene expression profiles given a third gene. Liquid association quantifies the evolving dependence structure between two univariate random variables by incorporating a third variable and measuring a three-way interaction. Extensions along this line of work have been developed over the years \citep{chen2011penalized, ho2011modeling, li2004system, yu2018new}. More recently, \citet{li2023generalized} introduced the generalized liquid association analysis for high-dimensional settings with three sets of continuous multivariate variables. However, existing methods are designed specifically for continuous conditioning variables within the three-way interaction framework. While treating the binary $Z$ variable as continuous is numerically feasible---for example, applying the penalized tensor decomposition algorithm in \citet{li2023generalized} with minimal modifications---such an approach leads to ambiguous or questionable interpretations. In particular, both the original liquid association framework \citep{li2002genome} and its generalized extension \citep{li2023generalized} quantify the expected derivative of the conditional association with respect to $Z$, implying smooth trends or continuous modulation. Binary variables, however, encode discrete group membership, and treating them as continuous can obscure group-based interpretations and lead to model misspecification. Our proposed conditional cross-covariance reduction (CCR) model and its estimation method are specifically designed for binary $Z$ and consequently provides a justified approach and new interpretation to the liquid association literature in contexts such as sex dimorphism in the TMJ.

To investigate sex differences in TMJ mechanics, we propose a CCR model, which provides a simple interpretation and quantification that naturally leads to estimation of sparse linear combinations of TMJ skull and muscle measurements that maximize differences in association by sex. Given the small sample size of the cadaver dataset, traditional cross-validation and penalization methods are unsuitable for selecting the most important variables. To address this, we develop a sequential permutation for selecting sparsity (SPSS) method as a stable data-driven approach for variable selection. 

This paper makes several contributions. First, we introduce an interpretable model for characterizing dynamic associations between two sets of variables conditioning on a third binary variable. Second, our estimation method incorporates variable selection through sparse singular value decomposition combined with hard thresholding, which helps reduce potential bias introduced by penalization. Third, the SPSS method enhances robust feature selection, particularly in small-sample settings. Finally, these methodological contributions help provide stable, interpretable, and biologically meaningful insights into sex-specific biomechanical variations in the TMJ.

The rest of the paper is organized as follows. Section \ref{M-sec:methodology} introduces the CCR model and estimation procedures with the SPSS method for variable selection. We numerically show the CCR results and accuracy of the SPSS method in Section \ref{M-sec:simulation} and illustrate the TMJ analysis results in Section \ref{M-sec:realdata}. We conclude with a brief discussion in Section \ref{M-sec:discussion}. Supplementary Materials include additional numerical results and extensions.

\section{METHODOLOGY}\label{M-sec:methodology}

\subsection{Conditional Cross-covariance Reduction Model}

We first introduce the concept of a conditional cross-covariance reduction (CCR) model and tailor it to the case of the binary third variable (e.g., sex). The CCR model accommodates both continuous and discrete third variables, but our development focuses on the discrete case to facilitate our goal of identifying sexual dimorphism. 

Let the two sets of random variables be $\mbX \in \mathbb{R}^{p_{1}}$ and $\mbY \in \mathbb{R}^{p_{2}}$, and the third random variable $Z \in \mathbb{R}$. Then the conditional cross-covariance, which summarizes important aspects of the relationship between $\mbX$ and $\mbY$, can be formulated as $\Cov(\mbX,\mbY \mid Z=z)\equiv \bolSigma_{\mbX\mbY}(z) \in \mathbb{R}^{p_{1} \times p_{2}}$. Our CCR model assumes that the matrix $\bolSigma_{\mbX\mbY}(z)$ varies within low-dimensional subspaces for all values of $z$ as follows:
\begin{equation}
    \bolSigma_{\mbX\mbY}(z)=\Cov(\mbX,\mbY\mid Z=z)=\bolGamma_{1} f(z) \bolGamma_{2}^{\top} \in \mathbb{R}^{p_{1}\times p_{2}},
\label{M-eq:CCR}
\end{equation}
for some semi-orthogonal basis matrices $\bolGamma_{1} \in \mathbb{R}^{p_{1} \times d_{1}}$ and $\bolGamma_{2} \in \mathbb{R}^{p_{2} \times d_{2}}$, and some latent function $f:\mathbb{R} \mapsto \mathbb{R}^{d_{1} \times d_{2}}$. The latent matrix-variate function $f(z) \in \mathbb{R}^{d_{1} \times d_{2}}$, where $d_{1} \leq p_{1}$ and $d_{2} \leq p_{2}$, is what drives the dynamic covariance between $\mbX$ and $\mbY$. We note that although the matrices $\bolGamma_{1}$ and $\bolGamma_{2}$ are not unique, the subspaces spanned by their column vectors are. 
Under the CCR model \eqref{M-eq:CCR}, the linear combinations $\bolGamma_{1}^{\top}\mbX$ and $\bolGamma_{2}^{\top}\mbY$ capture associations in $\mbX$ and $\mbY$ that vary with $z$. The function $f(\cdot)$ contains the coordinates of the conditional cross-covariance $\bolSigma_{\mbX \mbY}(z)$ relative to $\bolGamma_{1}$ and $\bolGamma_{2}$. Thus, the important signals in the rows and columns are preserved by $\mathrm{span}(\bolGamma_{1})$ and $\mathrm{span}(\bolGamma_{2})$, respectively. 

When the third variable is binary $Z \in \{1,2\}$, the CCR model implies that the variation in $\bolSigma_{\mbX\mbY}(z)$ along $z$ is fully characterized by the non-stochastic matrix $\bolPhi=\bolSigma_{\mbX\mbY}(1)-\bolSigma_{\mbX\mbY}(2)$. Furthermore, the singular value decomposition (SVD) of $\bolPhi$ implies that the dimensions of the latent subspaces have to be $d_{1}=d_{2}=r$ for some integer $r$. Then, we have $\bolPhi=\mbU\mbD\mbV^{\top}$ where $\mbU \in \mathbb{R}^{p_{1} \times r}$, $\mbV \in \mathbb{R}^{p_{2} \times r}$ are orthonormal basis matrices and $\mbD \in \mathbb{R}^{r \times r}$ is a diagonal matrix. The rank $r$ is a pre-specified value. In practice, we may take the rank as 1 or 2 for exploratory analysis and data visualization. The rank selection is still an open question in low-rank matrix approximation, with many ad-hoc approaches proposed in the matrix decomposition literature, and is beyond the scope of this paper. 

\subsection{Subspace Estimation}\label{M-sec:subspace_estimation}
In the CCR model, association patterns in $\mbX$ and $\mbY$ that are affected by $Z$ can be fully captured by linear combinations of $\mbU^{\top}\mbX$ and $\mbV^{\top}\mbY$. We estimate the subspace $\mathrm{span}(\mbU)$ spanned by the columns of $\mbU$ and the subspace $\mathrm{span}(\mbV)$ spanned by the columns of $\mbV$.
For $N$ i.i.d. observations $\{\mbx_{i},\mby_{i},z_{i}, i=1,\ldots,N \}$, let the first $n_{1}$ observations have $z_{i}=1$ that the remaining $n_{2}=N-n_{1}$ observations have $z_{i}=2$. We center the data within each group because we are interested in the conditional cross-covariance and not the conditional means. For simplicity, we assume that the data are already centered so that $\sum_{i=1}^{n_{1}}\mbx_{i}=\sum_{i=n_{1}+1}^{N}\mbx_{i}=\textbf{0}$, and $\sum_{i=1}^{n_{1}}\mby_{i}=\sum_{i=n_{1}+1}^{N}\mby_{i}=\textbf{0}$. We estimate the subspaces $\mbU$ and $\mbV$ as follows:
\begin{align}
(\widetilde{\mbU}, \widetilde{\mbV}) 
&=\argmax_{\mbU, \mbV} \Big\{ \widehat{\cov}(\mbU^{\top}\mbX, \mbV^{\top}\mbY \mid Z=1)-\widehat{\cov}(\mbU^{\top}\mbX, \mbV^{\top}\mbY \mid Z=2) \Big\} \nonumber \\
&=\argmax_{\mbU, \mbV}\| \mbU^{\top} \widehat{\bolSigma}_{\mbX\mbY1}\mbV-\mbU^{\top} \widehat{\bolSigma}_{\mbX\mbY2}\mbV\|^{2}_{\mathrm{F}},
\label{M-eq:arg_max}
\end{align}
where $\widehat{\bolSigma}_{\mbX\mbY1}=\frac{1}{n_{1}}\sum_{i=1}^{n_{1}}\mbx_{i}\mby_{i}^{\top}$ and $\widehat{\bolSigma}_{\mbX\mbY2}=\frac{1}{n_{2}}\sum_{i=n_{1}+1}^{N}\mbx_{i}\mby_{i}^{\top}$, and $\|\cdot\|_{\mathrm{F}}$ represents the Frobenius norm. 
Under orthogonality constraints, the resulting $(\widetilde{\mbU},\widetilde{\mbV})$ are the left and right singular vectors of $\widetilde{\bolPhi}=\frac{1}{n_{1}}\sum_{i=1}^{n_{1}}\mbx_{i}\mby_{i}^{\top}-\frac{1}{n_{2}}\sum_{i=n_{1}+1}^{N}\mbx_{i}\mby_{i}^{\top}$, provided that the SVD is well-defined (e.g., sufficient sample size to ensure a non-singular $\widetilde{\bolPhi}$). To enhance interpretability and also to deal with the very small sample size in our study, we next incorporate variable selection to further reduce the number of parameters. 
 
\subsection{Variable Selection and Algorithm}\label{M-sec:sparsity}

We consider sparsity on the singular vectors of $\bolPhi$, which is achievable by many existing sparse SVD algorithms such as the iterative thresholding algorithm in \citet{yang2016rate}. In our CCR model, we pre-specify the sparsity levels as $s_{1}\leq p_1$ and $s_{2}\leq p_2$ according to the elements in $\mbX$ and $\mbY$ that have dynamic association to each other instead of applying threshold tuning parameters iteratively in the sparse SVD algorithm. Thus, at each iteration, we keep $s_{1}$ and $s_{2}$ variables in $\mbX$ and $\mbY$. Then we get the singular values having the largest $r$ components and the corresponding singular vectors with $s_{1}$ and $s_{2}$ non-zero components. This sparse estimation performs variable selection for $\bolPhi$ since the estimated $\widehat{\bolPhi}$ has the $s_{1}$ and $s_{2}$ variables most strongly tied to the patterns of association in $\mbX$ and $\mbY$. Thus, we can effectively elucidate the associations between modalities that exhibit a maximal difference by sex, and, simultaneously, identify a sparse set of variables driving these associations.

We summarize the estimation procedure in Algorithm \ref{M-alg:BCCR via sparsity}, which yields $\widehat{\mbU}$ and $\widehat{\mbV}$, at the desirable input sparsity levels, $s_{1}$ and $s_{2}$.
The iteration is initialized with $\widehat{\mbU}^{(0)}$ and $\widehat{\mbV}^{(0)}$, the left and right orthonormal matrices of $\widetilde{\bolPhi}$. Each iteration updates these values by first computing multiplication forms $\widehat{\mbU}^{(t),\mathrm{mul}}$ and $\widehat{\mbV}^{(t), \mathrm{mul}}$ that extract the leading eigenvectors (steps (3a) and (3d)), then applying rowwise thresholding to enforce sparsity (steps (3b) and (3e)), yielding $\mbU^{(t),\mathrm{thr}}$ and $\mbV^{(t),\mathrm{thr}}$, and then orthonormalization (steps (3c) and (3f)) to update the subspace estimates.  Upon convergence, these retained rows represent the variables selected under the sparsity constraint that provide the linear combinations of $\mbX$ and $\mbY$ that are most contrastive for the values of $Z$. Convergence is determined by a tolerance on the maximum subspace difference: 
$\max\big(\|\widehat{\mbU}^{(t)}\widehat{\mbU}^{(t)\top}-\widehat{\mbU}^{(t-1)}\widehat{\mbU}^{(t-1)\top}\|^{2}_{\mathrm{F}} \enskip ,\|\widehat{\mbV}^{(t)}\widehat{\mbV}^{(t)\top}-\widehat{\mbV}^{(t-1)}\widehat{\mbV}^{(t-1)\top}\|^{2}_{\mathrm{F}} \big) \leq \epsilon$.

\begin{algorithm}[!htb]
\caption{CCR model via two-way iterative thresholding}
\begin{algorithmic}[1]
\Inputs{The sample estimate $\widetilde{\bolPhi} \in \mathbb{R}^{p_{1} \times p_{2}}$, the corresponding rank $r \leq \mathrm{min}(p_{1},p_{2})$, and the sparsity levels $s_{1} \leq p_{1}$, $s_{2} \leq p_2$.}
\medskip
\Initialize{Compute the top-$r$  singular vectors of $\widetilde{\bolPhi}$, $\widehat{\mbV}^{(0)}  \in \mathbb{R}^{p_2 \times r}$ and $\widehat{\mbU}^{(0)}  \in \mathbb{R}^{p_1 \times r}$.}
\State \textbf{Repeat $t=1,2,\ldots$}
\begin{enumerate}[label=(\alph*)]
\item Left multiplication: $\mbU^{(t),\text{mul}}=\widetilde{\bolPhi}\widehat{\mbV}^{(t-1)}.$
\item Left thresholding: for $I \subseteq \{1,2,\ldots,p_{1}\}$ and $i=1,\ldots,p_{1}$,
\begin{equation*}
    \mbU_{i}^{(t),\text{thr}}=
    \begin{cases}
        \mbU_{i}^{(t),\text{mul}} & , i \in \{\argmax_{|I|=s_{1}} \sum_{l \in I} \|\mbU_{l}^{(t),\text{mul}}\|_{2}\} \\
        0 & , \text{otherwise}
    \end{cases}
\end{equation*}

\item Left orthonormalization: QR decomposition on $\mbU^{(t),\text{thr}}$,
\begin{equation*}
\text{such that} \ \widehat{\mbU}^{(t)} \ \text{satisfies}\ \mathrm{span}(\widehat{\mbU}^{(t)})=\mathrm{span}(\widehat{\mbU}^{(t),\text{thr}}) \  \text{when} \ \{\widehat{\mbU}^{(t)}\}^{\top}\widehat{\mbU}^{(t)}=\mbI_{r}.     
\end{equation*}
\item Right multiplication:: $\mbV^{(t),\text{mul}}=\widetilde{\bolPhi}^{\top}\widehat{\mbU}^{(t)}.$
\item Right thresholding: for $J \subseteq \{1,2,\ldots,p_{2}\}$ and $j=1,\ldots,p_{2}$,
\begin{equation*}
    \mbV_{j}^{(t),\text{thr}}=
    \begin{cases}
        \mbV_{j}^{(t),\text{mul}} & , j \in \{\argmax_{|J|=s_{2}} \sum_{l \in J} \|\mbV_{l}^{(t),\text{mul}}\|_{2}\} \\
        0 & , \text{otherwise}
    \end{cases}
\end{equation*}

\item Right orthonormalization: QR decomposition on $\mbV^{(t),\text{thr}}$,
\begin{equation*}
\text{such that} \ \widehat{\mbV}^{(t)} \ \text{satisfies}\ \mathrm{span}(\widehat{\mbV}^{(t)})=\mathrm{span}(\widehat{\mbV}^{(t),\text{thr}}) \  \text{when} \ \{\widehat{\mbV}^{(t)}\}^{\top}\widehat{\mbV}^{(t)}=\mbI_{r}.     
\end{equation*}
\end{enumerate}
\textbf{until} convergence.
\Output{$\widehat{\mbU}=\widehat{\mbU}^{(t)}$, $\widehat{\mbV}=\widehat{\mbV}^{(t)}$, $\mbP_{\widehat{\mbU}}=\widehat{\mbU}^{(t)}\widehat{\mbU}^{(t)\top}, \mbP_{\widehat{\mbV}}=\widehat{\mbV}^{(t)}\widehat{\mbV}^{(t)\top}$, and $\widehat{\bolPhi}=\mbP_{\widehat{\mbU}}\widetilde{\bolPhi}\mbP_{\widehat{\mbV}}$.}

\end{algorithmic}
\label{M-alg:BCCR via sparsity}
\end{algorithm}

\subsection{Covariance and Correlation Differences}

We define the \textit{maximal covariance difference} $\delta_{i}=\mbU_{i}^{\top}\bolPhi\mbV_{i} >0$ where $\mbU_{i}$ and $\mbV_{i}$ are the $i$-th pair of the singular vectors of $\bolPhi$. When $r = 1$, $\delta_{1}$ is the maximal covariance difference that increases as we increase the sparsity parameters $s_{1}$ and $s_{2}$. More generally, $\delta_{i}$ can be defined for $i = 1, \ldots, r$ with rank $r$. We also define the \textit{associated correlation difference} $\eta_{i}$ as follows:
$$\eta_{i}=\Corr(\mbU_{i}^{\top}\mbX,\mbV_{i}^{\top}\mbY\mid Z=1)-\Corr(\mbU_{i}^{\top}\mbX,\mbV_{i}^{\top}\mbY\mid Z=2) \in \mathbb{R}, \ i=1,\ldots,r,$$
where by ``associated'' we mean that the vectors $\mbU_{i}$ and $\mbV_{i}$ are defined from the maximizing association differences (in covariance scales). It is difficult to simultaneously optimize the subspaces for the difference of two canonical correlation forms in terms of the associated correlation difference $\eta_{i}$. Thus, the correlation difference is calculated with the subspaces $\mbU$ and $\mbV$ that are estimated from the maximization problem in \eqref{M-eq:arg_max}. Even if we marginally standardize $\mbX$ and $\mbY$, we still maximize the marginally standardized form of $\delta_{i}$, not $\eta_{i}$. We use the associated correlation difference $\eta_{i}$ to demonstrate that the proposed CCR model avoids the masking of the association between $\mbX$ and $\mbY$ by $Z$.

\subsection{Sequential Permutation for Selecting Sparsity}\label{M-sec:selection}
 Algorithm \ref{M-alg:BCCR via sparsity} requires the sparsity levels $s_{1}$, $s_{2}$. In Supplementary Materials Section \ref{S-sec:CCR_IC}, we introduce an information criterion and illustrate its consistency in selecting $s_1$ and $s_2$, both theoretically when $N\rightarrow\infty$ and numerically with simulations. However, due to the limited size of the cadaver dataset, we find that either information criterion or cross-validation can select the sparsity levels accurately. Instead, we devise a sequential permutation for selecting sparsity (SPSS) approach to select $s_{1}$ and $s_{2}$ separately. We employ a leave-two-out (LTO) resampling scheme that iteratively removes one observation from each group $Z \in \{1,2\}$ and fits the CCR model to the remaining $N-2$ samples. The SPSS approach considers hypotheses related to the increment of the nuclear norm of $\widehat{\bolPhi}=\mbP_{\widehat{\mbU}}\widetilde{\bolPhi}\mbP_{\widehat{\mbV}}$ from the output of Algorithm \ref{M-alg:BCCR via sparsity}. When $r=1$, the nuclear norm of $\widehat{\bolPhi}$ reduces to $\widehat{\delta}_{1}$. 

Considering $s_{1}$, we sequentially postulate that $s_{1}=i$, $i=1,2,\ldots,p_{1}-1$, until $s_{1}=i+1$ does not improve upon $s_{1}=i$, at which point we take $s_{1}=i$. The larger value of $s_{1}=i+1$ fails to improve upon $s_{1}=i$ when at least one hypothesis $H_{0}^{i,k}: \overline{\delta}_{1}^{(i+1,k)}-\overline{\delta}_{1}^{(i,k)}=0$ is not rejected in favor of the one-sided alternative $H_{1}^{i,k}: \overline{\delta}_{1}^{(i+1,k)}-\overline{\delta}_{1}^{(i,k)}>0$, $k=1,2,\ldots,p_{2}$, where $\overline{\delta}_{1}^{(i,k)}$ is the population counterpart of the sample algorithm's output $\widehat{\delta}_{1}$ at the sparsity level $(s_{1},s_{2})=(i,k)$. 

The hypothesis test $H_{0}^{i,k}$ vs $H_{1}^{i,k}$ is performed using a permutation procedure. Let $\widetilde{\delta}_{1}^{(i,k)}$ be a sample counterpart of $\overline{\delta}_{1}^{(i,k)}$, and define $D_{\ell}$, $\ell=1,\ldots,(n_{1} n_{2})$, as the differences $\widehat{\delta}_{1}^{(i+1,k)}-\widehat{\delta}_{1}^{(i,k)}$ from $(n_{1} n_{2})$ LTO data splits. Then the observed mean difference is $\widetilde{\delta}^{(i+1,k)}_{1}-\widetilde{\delta}^{(i,k)}_{1} = (n_{1}n_{2})^{-1} \sum_{\ell=1}^{n_{1}n_{2}} D_{\ell}$. To obtain a reference distribution under the null hypothesis, we use a large number of permutations—for example, 100,000 in our TMJ analysis in Section \ref{M-sec:realdata}—where, in each permutation, the signs of the $D_{\ell}$ values are randomly flipped before averaging. The p-value, denoted $p^{(i,k)}$, is the proportion of permuted means at least as large as the observed mean difference $\widetilde{\delta}^{(i+1,k)}_{1}-\widetilde{\delta}^{(i,k)}_{1}$, which serves as the test statistic. Thus $s_{1}$ is set to the smallest $i$ such that the collection of p-values $p^{(i,k)}$, $k=1,\ldots,p_{2}$, has at least one value larger than 0.05. An analogous procedure is used to determine $s_{2}$. 

 \section{SIMULATION}\label{M-sec:simulation}

 \subsection{Simulation Setup}\label{M-sec:sim_setup}
We perform simulations to examine the empirical performance of the proposed CCR model with a binary variable $Z$. We consider two scenarios for the rank of the cross-covariance, $r=1$ and $r=2$. We first set the rank of the cross-covariance of $\mbX$ and $\mbY$ as $r=1$, $p_1=18$, $p_2=15$, and true sparsities $s_{1}^{*}=s_{2}^{*}=3$, and vary the sample sizes $N=n_{1}+n_{2}$ from 40 to 400 when $n_1=n_2$. Under the rank-1 scenario for $\bolPhi$, i.e., $r=1$, we get $\bolPhi \bolPhi^{\top}=(\rho_{1}-\rho_{2})^{2}$ where $\rho_{1}-\rho_{2}$ is a coefficient of the SVD structure of $\bolPhi$. That is, maximizing $\bolPhi \bolPhi^{\top}$ is equivalent to maximizing the difference $\rho_{1}-\rho_{2}$ restricted to this difference being positive. 

We generate the data in the following way. For $i=1,\ldots,n_1$, we generate $(\mbx_{i},\mby_{i})$ jointly from a normal distribution with mean zero and covariance $\bolSigma_{1}$. For $i=n_{1}+1,\ldots,N$, we generate $(\mbx_{i},\mby_{i})$ jointly from a normal distribution with mean zero and covariance $\bolSigma_{2}$. Here, 
\begin{align}
\bolSigma_{z}=\begin{pmatrix}
\bolSigma_{\mbX} & \rho_{z}\mbU\mbV^{\top} \\
\rho_{z}\mbV\mbU^{\top} & \bolSigma_{\mbY}	
\end{pmatrix}, \enskip z=1,2,
\label{M-eq:covariance_structure}
\end{align}
where the group index $z$ represents the binary variable $Z \in \{1,2\}$. To maintain the positive-definiteness of the full covariance matrix and rank-1 condition for the cross-covariance of $\mbX$ and $\mbY$, we set $\mbU=\bolSigma_{\mbX}^{1/2}\mbO_{1}$ and $\mbV=\bolSigma_{\mbY}^{1/2}\mbO_{2}$ where the $\mbO_{1}$ and $\mbO_{2}$ are unit length vectors $\mbO_{1}=(1,1,1,0,\ldots,0)^{\top}/\sqrt{3} \in \mathbb{R}^{18 \times 1} \enskip \text{and}\enskip \mbO_{2}=(1,1,1,0,\ldots,0)^{\top}/\sqrt{3} \in \mathbb{R}^{15 \times 1}$. The columns of $\mbU$ and $\mbV$ are the subspace capturing the variation caused by $\mbX$ and $\mbY$, respectively. The CCR model $\bolPhi$ is defined as follows:
\begin{equation}
 \bolPhi=(\rho_1-\rho_2)\mbU\mbV^{\top}, 
 \label{M-eq:population_parameter}	
 \end{equation}
 where $\rho_{1}-\rho_{2}>0$. For the covariance matrix $\bolSigma_{z}$, the marginal covariance matrix $\bolSigma_{\mbX}$ is set as a block diagonal matrix, $\bolSigma_{\mbX}=\text{bdiag}(c_1\bolSigma_{\mbX,1},c_2\bolSigma_{\mbX,2})$, where $\bolSigma_{\mbX,1} \in \mathbb{R}^{s_{1}^{*} \times s_{1}^{*}}$ corresponds to non-zero elements and takes the form of an autoregressive (AR) structure such that its $(i,j)$th entry equals $\sigma_{ij}={0.7}^{|i-j|},\enskip i,j=1,\ldots,s_{1}^{*}$, and $\bolSigma_{\mbX,2}\in \mathbb{R}^{(p_{1}-s_{1}^{*})\times(p_{1}-s_{1}^{*})}$ is the identity matrix. The marginal covariance matrix $\bolSigma_{\mbY}$ is constructed similarly. Therefore, the true signals in $\bolSigma_{\mbX,1}$ and $\bolSigma_{\mbY,1}$ will be captured if our algorithm works correctly. We set $\rho_1=0.9$, $\rho_2=-0.9$, $c_{1}=3$, and $c_{2}=1$. Note that the larger the ratio $c_{1}/c_{2}$, the easier it is to detect the true signals.

For rank-2 simulation scenario ($r=2$), we modify the cross-covariance in \eqref{M-eq:covariance_structure} as $\mbU \mbD  \mbV^{\top}$, $z=1,2$, where $\mbD=\mathrm{diag}(\rho_{z1}, \rho_{z2})$, $\mbU=\bolSigma_{\mbX}^{1/2}\mbO_{1}$, and $\mbV=\bolSigma_{\mbY}^{1/2}\mbO_{2}$. Here, $\mbO_{1} \in \mathbb{R}^{p_{1} \times 2}$ and $\mbO_{2} \in \mathbb{R}^{p_{2} \times 2}$, with the first column being $(1,1,1,0,\ldots,0)/\sqrt{3}$, and the second column being $(0,-1,1,0,\ldots,0)/\sqrt{2}$ under the true sparsity levels $(s_{1}^{*},s_{2}^{*})=(3,3)$. For the rank-2 scenario, we set $p_{1}=18$, $p_{2}=15$, $\rho_{11}=0.9$, $\rho_{12}=0.7$, $\rho_{21}=-0.9$, $\rho_{22}=-0.7$, $c_{1}=3$, $c_{2}=1$, and change the sample size $N=n_{1}+n_{2}$ from 40 to 400 when $n_{1}=n_{2}$. There are two contributing linear combinations on each of $\mbX$ and $\mbY$ since $\mbU \in \mathbb{R}^{p_{1} \times 2}$ and $\mbV \in \mathbb{R}^{p_{2} \times 2}$. 

To evaluate the performance of each method in terms of variable selection and subspace estimation accuracy, we used a true positive rate (TPR), a false positive rate (FPR), and a subspace distance. We record the TPR and FPR for each row (variables selected from $\mbX$) and column (variables selected from $\mbY$) to assess the sparsity assumptions. Let $\calI_{\mbU} \subseteq \{1,2,\ldots,p_{1}\}$ the set of true nonzero rows of $\mbU$. The estimated index set is $\calI_{\widehat{\mbU}}=\{i:\text{there exist non-zero elements on the $i$th row of $\widehat{\mbU}$}\}$. Then the TPR is defined as the proportion of correctly selected variables, $\text{TPR}_{\mbX}=|\calI_{\mbU} \cap \calI_{\widehat{\mbU}}|/s_{1}^{*}$, and the FPR is the proportion of falsely selected variables, $\text{FPR}_{\mbX}=|\calI_{\mbU}^{c} \cap \calI_{\widehat{\mbU}}|/(p_{1}-s_{1}^{*})$. The definitions for $\text{TPR}_{\mbY}$ and $\text{FPR}_{\mbY}$ follow analogously by replacing $\mbU$, $\widehat{\mbU}$, $p_{1}$, $s_{1}^{*}$ with $\mbV$, $\widehat{\mbV}$, $p_{2}$, $s_{2}^{*}$. Next, for the estimation of $\mbU$ and $\mbV$, we compute the subspace distance between the true and estimated $\mbU$ and $\mbV$, $D_{\mbU}=\|\mbP_{\mbU}-\mbP_{\widehat{\mbU}}\|_{\mathrm{F}}/\sqrt{2r}$ and $D_{\mbV}=\|\mbP_{\mbV}-\mbP_{\widehat{\mbV}}\|_{\mathrm{F}}/\sqrt{2r}$ where $r$ is the true rank of the cross-covariance and $\|\cdot\|_{\mathrm{F}}$ represents the Frobenius norm. We calculate the associated correlation difference between the first linear combinations as $\widehat{\eta}_{1}$ and between the second linear combinations as $\widehat{\eta}_{2}$.  

\subsection{Simulation Results}\label{M-sec:sim_result}

We first show how the CCR model captures the true signals of the sparsity levels $s_{1}$ and $s_{2}$. In Algorithm \ref{M-alg:BCCR via sparsity}, we set the tolerance level $\epsilon$ as $10^{-11}$ and provide the correct values of $s_{1}$ and $s_{2}$ as inputs. For estimation assessments ($\text{D}_{\mbU}$ and $\text{D}_{\mbV}$) and variable selection results (TPR and FPR), we use $\widehat{r}=r$. To examine the empirical differences of signals at each rank, we compute covariance and correlation differences when $\widehat{r}=2$, as summarized in Table \ref{M-table:diff_n}. Note that the true rank of the underlying model, denoted by $r$, is used for data generation in the simulation. In contrast, our algorithm operates with an estimated rank $\widehat{r}$, which may differ from $r$. We do not attempt to estimate the true rank $r$, as its identification remains an open problem in dimension reduction. Table \ref{M-table:diff_n} summarizes the simulation results in 100 replications for the two rank scenarios. We change the sample sizes in each group $n_{1}=n_{2} \in \{20,30,50,100,200\}$. The values of the TPR and FPR support that the CCR model accurately selects the variables that produce the most contrastive linear combination by the binary variable. The smaller subspace distances indicate that the CCR model accurately estimates the subspaces $\mbU$ and $\mbV$. Under the rank-1 scenario ($r=1$), the estimated $\widehat{\delta}_{2}$ has smaller values (than $\widehat{\delta}_{2}$ under rank-2 scenario) and converges to zero with increasing sample size, since the true signals are only in the first canonical directions. Thus, the values of $\widehat{\delta}_{2}$ under a rank-1 scenario indicate that the second canonical directions are not needed to capture the contrastive difference by $Z$. However, under the rank-2 scenario, $\widehat{\delta}_{2}$ increases with larger sample sizes and indicates that true signals exist in the second canonical directions under the rank-2 scenario. Furthermore, in both scenarios, the first associated correlation difference ($\widehat{\eta}_{1}$) is greater than $\widehat{\eta}_{2}$. The result indicates that the first linear combination captures the most contrastive pattern in $\mbX$ and $\mbY$.

\begin{table}[H]
\centering
\caption{Numerical evaluations under rank-1 and rank-2 scenarios for the cross-covariances over 100 data replicates. The numbers in parentheses report the standard error of the subspace distances $\text{D}_{\mbU}$, $\text{D}_{\mbV}$, covariance differences $\widehat{\delta}_{1}, \widehat{\delta}_{2}$, and the associated correlation differences $\widehat{\eta}_{1}$, $\widehat{\eta}_{2}$. Here, we used the true sparsity levels ($s_{1}^{*}=s_{1}$, $s_{2}^{*}=s_{2}$) for estimation.}
\begin{tabular}{cccccc}
\multirow{2}{*}{\begin{tabular}[c]{@{}c@{}}Rank\\ ($r=1$)\end{tabular}} & \multicolumn{5}{c}{$n_{1}=n_{2}$} \\  
& 20   & 30   & 50   & 100   & 200   \\ \hline 
\rowcolor{Gray}
$\text{TPR}_{\mbX}$ & 1.000 & 1.000 & 1.000 & 1.000 & 1.000 \\
$\text{TPR}_{\mbY}$ & 1.000 & 1.000 & 1.000 & 1.000 & 1.000 \\
\rowcolor{Gray}
$\text{FPR}_{\mbX}$ & 0.000 & 0.000 & 0.000 & 0.000 & 0.000 \\
$\text{FPR}_{\mbY}$ & 0.000 & 0.000 & 0.000 & 0.000 & 0.000 \\
\rowcolor{Gray}
$\text{D}_{\mbU}$   & \begin{tabular}[c]{@{}c@{}}0.085\\(0.000)\end{tabular} & \begin{tabular}[c]{@{}c@{}}0.069\\(0.000)\end{tabular} & \begin{tabular}[c]{@{}c@{}}0.060\\(0.000)\end{tabular} & \begin{tabular}[c]{@{}c@{}}0.038\\(0.000)\end{tabular} & \begin{tabular}[c]{@{}c@{}}0.027\\(0.000)\end{tabular} \\
$\text{D}_{\mbV}$   & \begin{tabular}[c]{@{}c@{}}0.113\\(0.001)\end{tabular} & \begin{tabular}[c]{@{}c@{}}0.084\\(0.000)\end{tabular} & \begin{tabular}[c]{@{}c@{}}0.075\\(0.000)\end{tabular} & \begin{tabular}[c]{@{}c@{}}0.049\\(0.000)\end{tabular} & \begin{tabular}[c]{@{}c@{}}0.037\\(0.000)\end{tabular} \\ 

\rowcolor{Gray}
$\widehat{\delta}_{1}$   & \begin{tabular}[c]{@{}c@{}}11.765\\(0.254)\end{tabular} & \begin{tabular}[c]{@{}c@{}}11.825\\(0.239)\end{tabular} & \begin{tabular}[c]{@{}c@{}}12.131\\(0.186)\end{tabular} & \begin{tabular}[c]{@{}c@{}}12.068\\(0.141)\end{tabular} & \begin{tabular}[c]{@{}c@{}}12.035\\(0.082)\end{tabular} \\
$\widehat{\delta}_{2}$   & \begin{tabular}[c]{@{}c@{}}0.635\\(0.028)\end{tabular} & \begin{tabular}[c]{@{}c@{}}0.514\\(0.019)\end{tabular} & \begin{tabular}[c]{@{}c@{}}0.388\\(0.015)\end{tabular} & \begin{tabular}[c]{@{}c@{}}0.285\\(0.012)\end{tabular} & \begin{tabular}[c]{@{}c@{}}0.184\\(0.008)\end{tabular} \\ 

\rowcolor{Gray}
$\widehat{\eta}_{1}$   & \begin{tabular}[c]{@{}c@{}}1.785\\(0.006)\end{tabular} & \begin{tabular}[c]{@{}c@{}}1.778\\(0.006)\end{tabular} & \begin{tabular}[c]{@{}c@{}}1.792\\(0.004)\end{tabular} & \begin{tabular}[c]{@{}c@{}}1.794\\(0.003)\end{tabular} & \begin{tabular}[c]{@{}c@{}}1.793\\(0.002)\end{tabular} \\
$\widehat{\eta}_{2}$   & \begin{tabular}[c]{@{}c@{}}0.533\\(0.019)\end{tabular} & \begin{tabular}[c]{@{}c@{}}0.424\\(0.016)\end{tabular} & \begin{tabular}[c]{@{}c@{}}0.323\\(0.011)\end{tabular} & \begin{tabular}[c]{@{}c@{}}0.239\\(0.009)\end{tabular} & \begin{tabular}[c]{@{}c@{}}0.150\\(0.006)\end{tabular} \\

\hline \hline 
\multirow{2}{*}{\begin{tabular}[c]{@{}c@{}}Rank\\ ($r=2$)\end{tabular}} & \multicolumn{5}{c}{$n_{1}=n_{2}$} \\  
& 20   & 30   & 50   & 100   & 200   \\ \hline 
\rowcolor{Gray}
$\text{TPR}_{\mbX}$ & 1.000 & 1.000 & 1.000 & 1.000 & 1.000 \\ 
$\text{TPR}_{\mbY}$ & 1.000 & 1.000 & 1.000 & 1.000 & 1.000 \\ 
\rowcolor{Gray}
$\text{FPR}_{\mbX}$ & 0.000 & 0.000 & 0.000 & 0.000 & 0.000 \\ 
$\text{FPR}_{\mbY}$ & 0.000 & 0.000 & 0.000 & 0.000 & 0.000 \\ 
\rowcolor{Gray}
$\text{D}_{\mbU}$   & \begin{tabular}[c]{@{}c@{}}0.197\\(0.013)\end{tabular} & \begin{tabular}[c]{@{}c@{}}0.177\\(0.011)\end{tabular} & \begin{tabular}[c]{@{}c@{}}0.112\\(0.007)\end{tabular} & \begin{tabular}[c]{@{}c@{}}0.088\\(0.006)\end{tabular} & \begin{tabular}[c]{@{}c@{}}0.057\\(0.004)\end{tabular} \\
\multirow{2}{*}{$\text{D}_{\mbV}$} & 0.190 & 0.141 & 0.117 & 0.082 & 0.058 \\ 
& (0.014) & (0.012) & (0.008) & (0.005) & (0.003) \\  
\rowcolor{Gray}
$\widehat{\delta}_{1}$   & \begin{tabular}[c]{@{}c@{}}11.772\\(0.248)\end{tabular} & \begin{tabular}[c]{@{}c@{}}11.861\\(0.239)\end{tabular} & \begin{tabular}[c]{@{}c@{}}12.151\\(0.187)\end{tabular} & \begin{tabular}[c]{@{}c@{}}12.079\\(0.142)\end{tabular} & \begin{tabular}[c]{@{}c@{}}12.055\\(0.082)\end{tabular} \\
$\widehat{\delta}_{2}$   & \begin{tabular}[c]{@{}c@{}}1.127\\(0.078)\end{tabular} & \begin{tabular}[c]{@{}c@{}}1.224\\(0.050)\end{tabular} & \begin{tabular}[c]{@{}c@{}}1.269\\(0.025)\end{tabular} & \begin{tabular}[c]{@{}c@{}}1.238\\(0.017)\end{tabular} & \begin{tabular}[c]{@{}c@{}}1.253\\(0.014)\end{tabular} \\

\rowcolor{Gray}
$\widehat{\eta}_{1}$   & \begin{tabular}[c]{@{}c@{}}1.791\\(0.006)\end{tabular} & \begin{tabular}[c]{@{}c@{}}1.787\\(0.006)\end{tabular} & \begin{tabular}[c]{@{}c@{}}1.796\\(0.004)\end{tabular} & \begin{tabular}[c]{@{}c@{}}1.794\\(0.003)\end{tabular} & \begin{tabular}[c]{@{}c@{}}1.796\\(0.002)\end{tabular} \\
\multirow{2}{*}{$\widehat{\eta}_{2}$} & 1.057 & 1.174 & 1.200 & 1.179 & 1.189 \\ 
& (0.060) & (0.036) & (0.014) & (0.009) & (0.006) 
\end{tabular}
\label{M-table:diff_n}
\end{table}
\normalsize

We also investigate empirical values of the maximal covariance differences under the rank-1 scenario. To specify the maximum covariance difference by the true sparsity levels, we use two different sparsity levels $\{(s_{1}^{*},s_{2}^{*})\}=\{(3,3),(10,10)\}$ and set $n_{1}=11$, $n_{2}=10$, $p_{1}=18$, and $p_{2}=15$. 

\begin{figure}[ht]
\centering
         \includegraphics[width=0.6\linewidth]{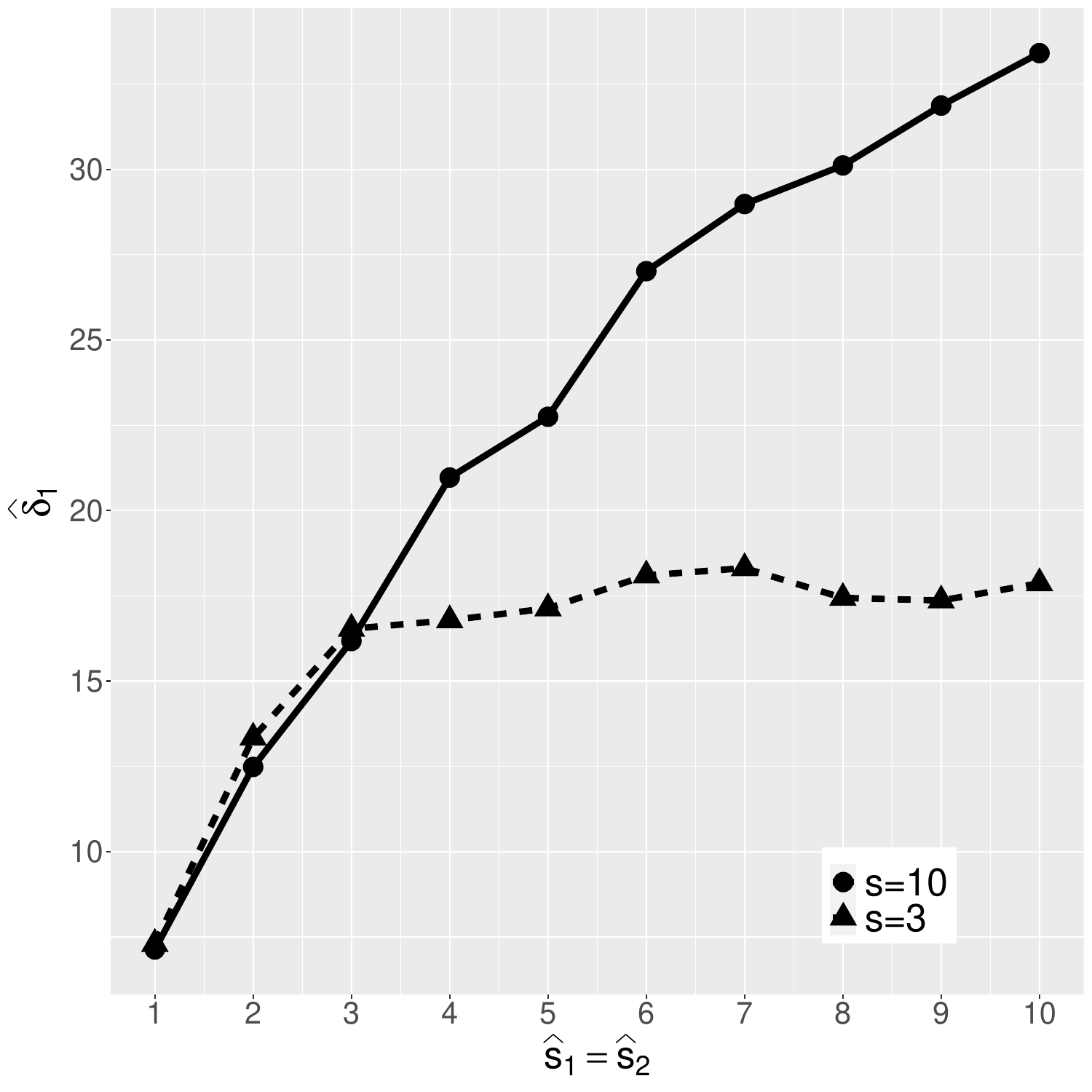}  	
\caption{Estimated maximal covariance difference ($\widehat{\delta}_{1}$) where $n_1=11$, $n_2=10$, $p_1=18$, $p_2=15$. In solid line with circle dots, the true signals are in the first ten rows and columns ($s_{1}^{*}=s_{2}^{*}=s=10$) and increase $s_{1}=s_{2}\in \{1,2,\ldots,10\}$. In dashed line with triangular dots, the true signals are in the first three rows and columns ($s_{1}^{*}=s_{2}^{*}=s=3$) and increase $s_{1}=s_{2}\in \{1,2,\ldots,10\}$.}
\label{M-fig:cor_cov_diff2}
\end{figure}

One would expect the maximal covariance differences to increase monotonically as the sparsity levels specified in Algorithm \ref{M-alg:BCCR via sparsity} increase because the information available for maximizing $\delta_1$ increases. In Figure \ref{M-fig:cor_cov_diff2}, this is indeed the case when the true sparsity levels are 10 ($s_{1}^{*}=s_{2}^{*}=10$) and the sparsity levels vary from 1 to 10 ($s_{1}=s_{2} = 1,2,\ldots,10$), represented by the solid line and circle dots. In this case, the estimated $\widehat{\delta}_{1}$ gradually increases. In contrast, when the true sparsity levels are 3 (represented by the dashed and triangular dots), $\widehat{\delta}_{1}$ rapidly increases as the sparsity levels ($s_{1}=s_{2}$) rise to 3, beyond which no substantial increase is observed. This rapid stabilization suggests that accurately estimating the true sparsity level is crucial, as it substantially impacts both the sensitivity and robustness of the maximal covariance difference. These distinct patterns highlight the CCR model’s ability to detect meaningful associations while avoiding overfitting, which is achieved by selecting appropriate sparsity levels through the SPSS method.

We conclude that the SPSS approach accurately identifies the sparsity levels of $\mbX$ and $\mbY$ when $c_{1}/c_{2} \geq 5$. The effect of the signal-to-noise ratio $c_{1}/c_{2}$ on SPSS, along with additional simulation scenarios—including an example with a multi-categorical conditioning variable $Z$—are provided in the Supplementary Materials.

 \section{REAL DATA ANALYSIS}\label{M-sec:realdata}

Our goal is to identify the variables associated with sexual dimorphism in the association between features of the TMJ temporalis origin muscle attachment and features of the skull. We set the skull measurement $\mbX \in \mathbb{R}^{16}$, the temporalis origin (TO) measurements in $\mbY \in \mathbb{R}^{18}$, and sex is the binary variable $Z \in \{1,2\}=\{\text{male},\text{female}\}$. We center each variable within sex group as discussed in Section \ref{M-sec:subspace_estimation}.

\begin{figure}[!ht]
\centering
         \includegraphics[width=\linewidth]{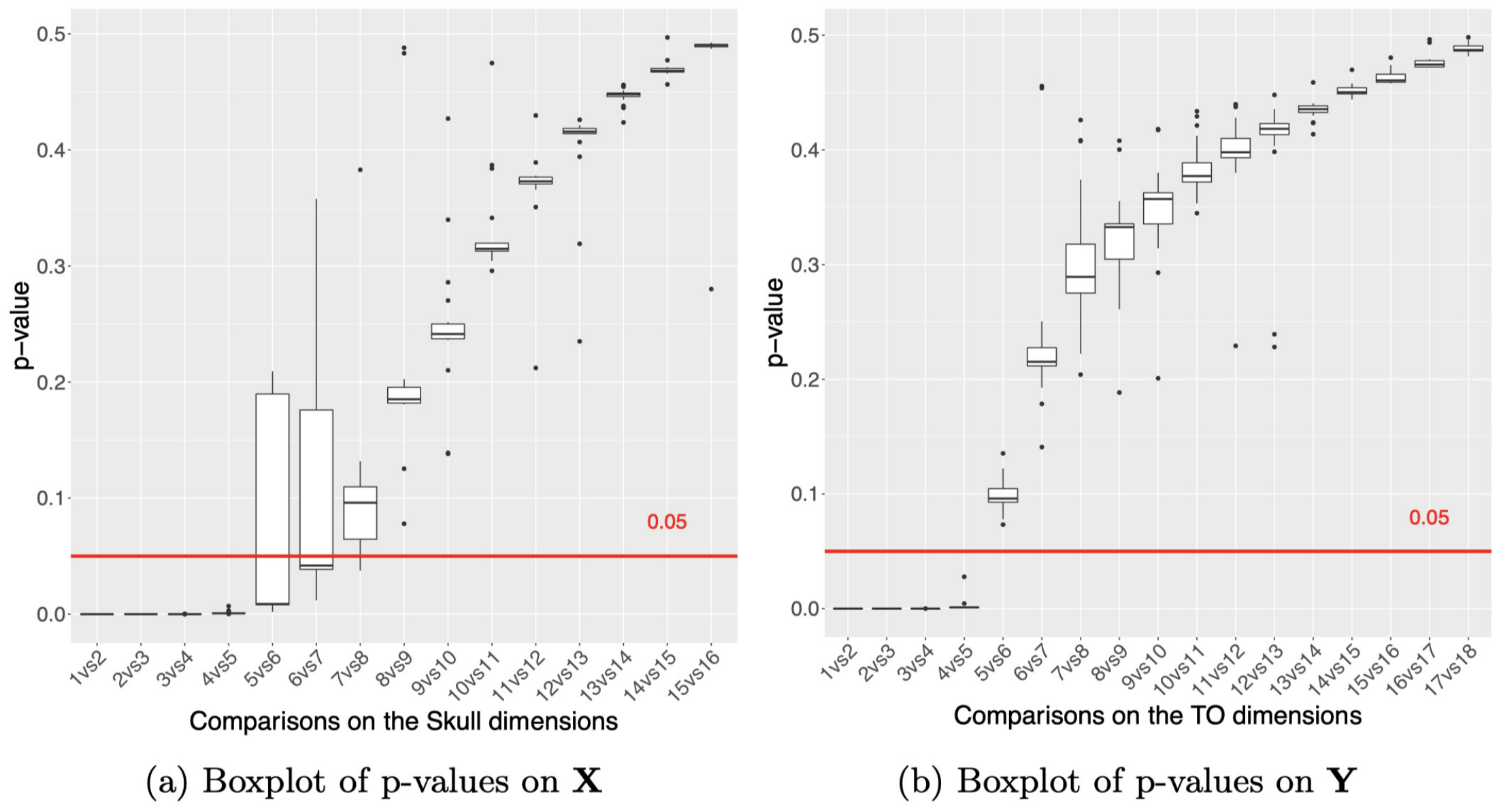}  	
         \caption{Box plots of p-values on the result of the sequential permutation for selecting sparsity (SPSS) method for the maximal covariance difference $\widehat{\delta}_{1}$ on skull and temporal origin (TO). From ``1vs2'', we can select the sparsity level $s_{1}$ and $s_{2}$ on panels (a) and (b), respectively. In panel (a), for example, all small p-values on ``1vs2'' ($<0.05$) represent that there are significant difference between $s_{1}=1$ and $s_{1}=2$ with fixed $s_{2}=k \in \{1,\ldots 18\}$. Then, we consecutively move to the next boxplots until there exists any p-value greater than 0.05 (no significant difference). Here, we select the sparsity level $(s_{1},s_{2})=(5,5)$.}
\label{M-fig:permtest}
\end{figure}

First, we investigate the sparsity levels for the CCR using the SPSS method described in Section \ref{M-sec:selection}. Figure \ref{M-fig:permtest} shows boxplots of the p-values $\{p^{(i,k)}\}$ for $i=1,\ldots,16$ and $k=1,\ldots,18$. The figure indicates that, for both $s_{1}$ and $s_{2}$, there is no significant difference between the sparsity levels of 5 and 6. Hence, we take $(s_{1},s_{2})=(5,5)$ in our analysis.

\begin{figure}[ht]
\centering
         \includegraphics[width=0.6\linewidth]{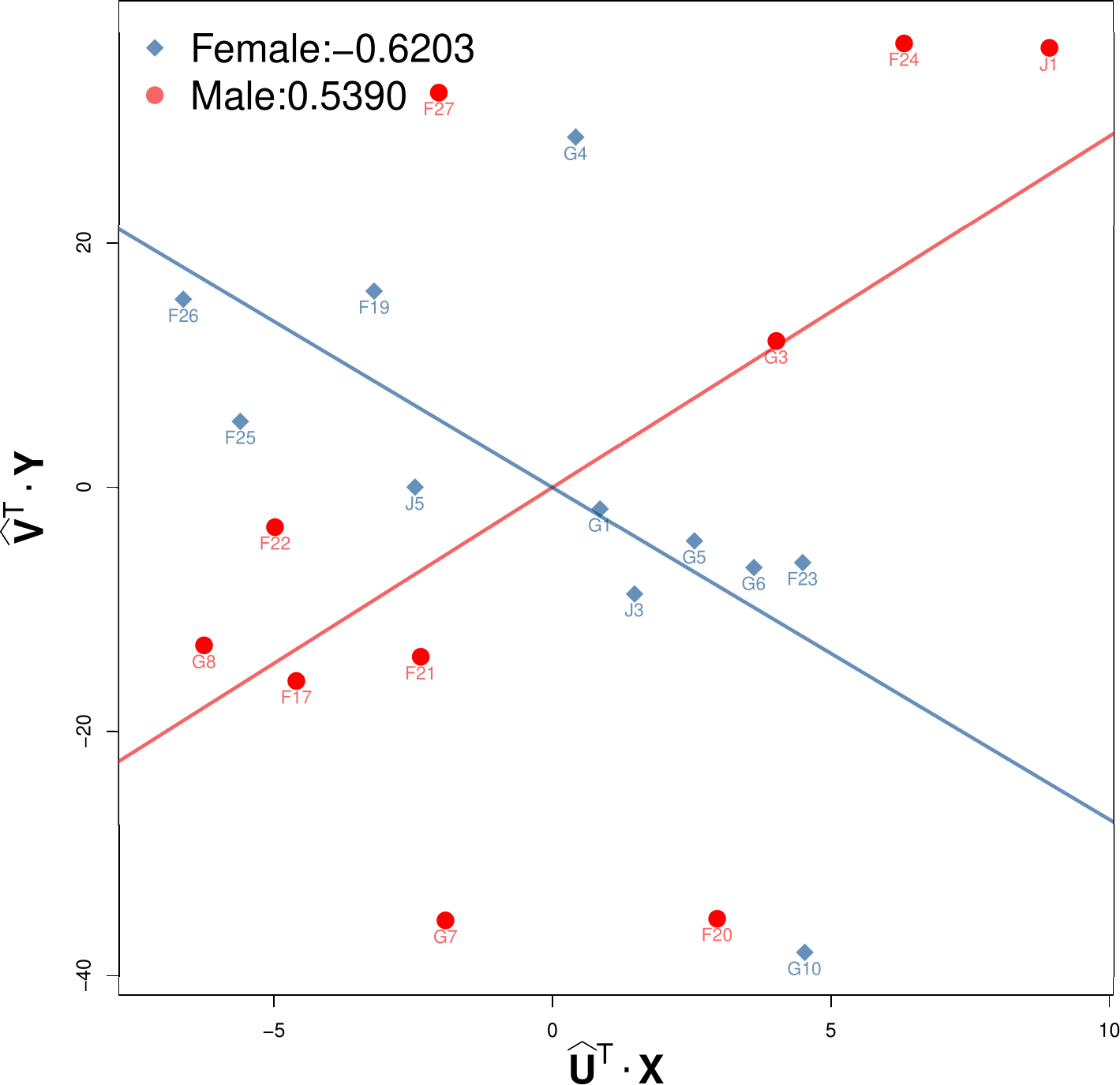}
\caption{Linear combinations of the CCR model on the skull and temporalis origin (TO) measurements. The numbers in the legend represent the correlations between linear combinations by sex. Sparsity levels are set as $s_{1}=s_{2}=5$.  The $x$-axis ($\widehat{\mbU}^{\top}\mbX$) indicates the linear combination on the skull, and the $y$-axis ($\widehat{\mbV}^{\top}\mbY$) represents the linear combination on temporalis origin (TO).}         
\label{M-fig:lin_comb_skull_vs_TO}
\end{figure}

Figure \ref{M-fig:lin_comb_skull_vs_TO} shows the linear combinations discovered by the CCR model from the skull ($\mbX$) and TO attachment ($\mbY$) measurements that have association that differs most by sex. The maximal covariance difference is $\widehat{\delta}_{1}=119.65$. The plot also displays correlations between the resulting skull and muscle attachment linear combinations by sex. The associated correlation difference $\widehat{\eta}_{1}=1.16$, which is the difference of the displayed correlations, demonstrates that the estimated subspaces differentiate association by sex. 

\begin{figure}[!ht]
\centering
\includegraphics[width=0.9\textwidth]{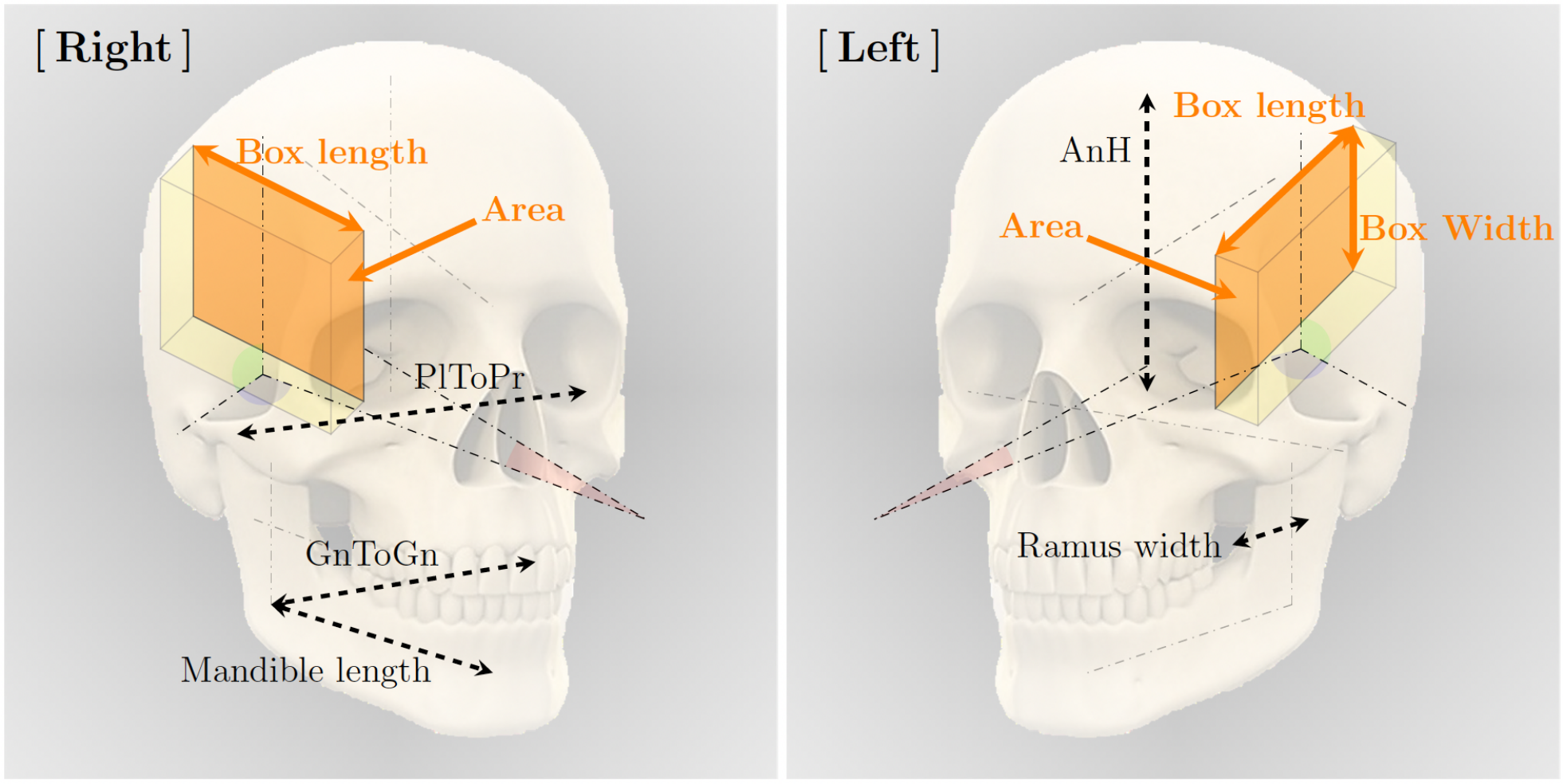}
\caption{Result of the CCR model on the skull and temporalis origin (TO) under $(s_{1},s_{2})=(5,5)$ where selected 10 variables are denoted with variable names. Selected variables on TO are marked as orange-colored bolded letters with solid lines, and selected variables on the skull are distinct as non-bolded letters with dashed lines.}
\label{M-fig:CCR_TMJ_result}
\end{figure}

The selected variables are in \eqref{M-eq:linear_combination_TMJ} and the corresponding skull images are shown in Figure \ref{M-fig:CCR_TMJ_result}. The selected variables in skull align with previous forensic anthropological results that state that the width of the bicondylar (PlToPr) and the width of the bigonial (GnToGn) are significant in determining the difference between the sexes \citep{de2016mandibular}. The bicondylar width (PlToPr) and the bigonial width (GnToGn) are involved in the dimensions of the medial lateral skull, and the length of the right side of the mandible is related to the size of the anterior and posterior skulls. All variables selected in the TO are related to the muscle attachment size (orange-colored bolded variables in Figure \ref{M-fig:CCR_TMJ_result}).

Recall that all data are centered within sex, and the linear combinations ($\widehat{\mbU}^{\top}\mbX$,$\widehat{\mbV}^{\top}\mbY$) are positively correlated with males (0.5390) and negatively correlated with females (-0.6203). Thus, within females, larger values of $\widehat{\mbU}^{\top}\mbX$ correspond to smaller values of $\widehat{\mbV}^{\top}\mbY$. One way to increase $\widehat{\mbU}^{\top}\mbX$ is through a smaller mandibular length (MandibleLength(R)), holding all else fixed. Therefore, a female with a smaller mandibular length (relative to other females), which increases $\widehat{\mbU}^{\top}\mbX$, corresponds to a smaller $\widehat{\mbV}^{\top}\mbY$. That is, a female with a smaller-than-average mandibular length (among females) may have a larger-than-average TO attachment size (among females). However, this association is not evident in raw one-to-one variable relationships, where smaller mandibular length corresponds to smaller attachment areas. The CCR model thus uncovers associations between selected variables that are masked in simple pairwise comparisons.

\begin{align}
    \text{Skull:} \enspace \widehat{\mbU}^{\top}\mbX &= 0.739\enspace \text{GnToGn}+0.308\enspace \text{AnH}+0.368\enspace \text{PlToPr} \label{M-eq:linear_combination_TMJ}\\
    &-0.323\enspace \text{RamusWidth(L)}-0.345\enspace \text{MandibleLength(R)}\nonumber  \\
    \text{TO:} \enspace \widehat{\mbV}^{\top}\mbY &=-0.462\enspace \text{BoxLength(L)}-0.482\enspace \text{BoxLength(R)}\nonumber  \\
    &-0.323\enspace \text{BoxWidth(L)}-0.524\enspace \text{Area(L)}-0.319\enspace \text{Area(R)} \nonumber
\end{align}

\begin{figure}
    \centering
\includegraphics[width=\linewidth]{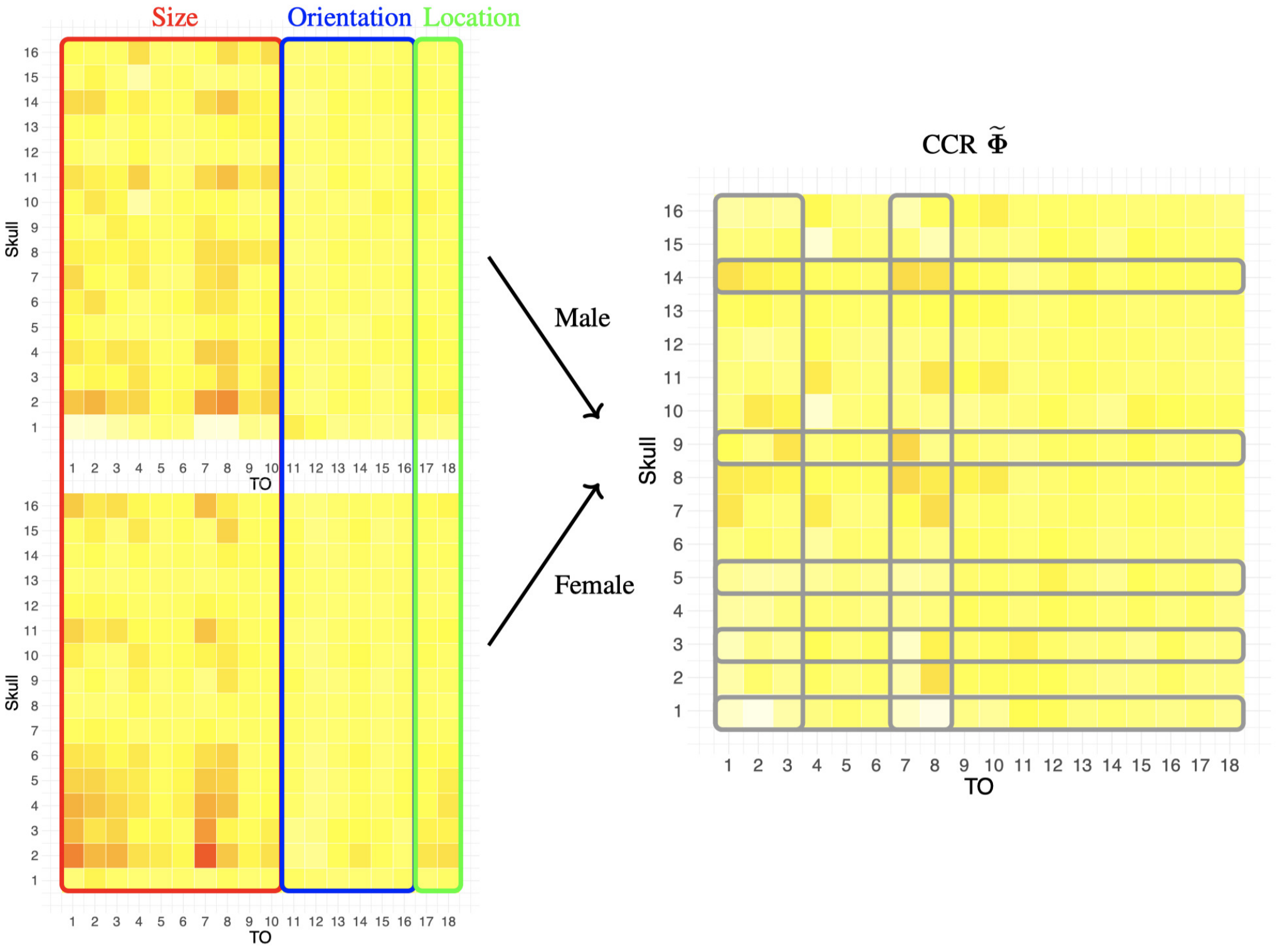}
\caption{Marginal covariance of male and female (left), and sample covariance difference $\widetilde{\bolPhi}$ (right). The variables subsetted in size, orientation, and location are displayed in Table \ref{S-table:variables} in Supplementary Materials. The gray boxes on the right side represent the variables selected under $(s_{1},s_{2})=(5,5)$.}
    \label{M-fig:TMJ_heatmap}
\end{figure}

Figure \ref{M-fig:TMJ_heatmap} shows the heat maps of the marginal covariances and $\widetilde{\bolPhi}$. The selected variables in $\widetilde{\bolPhi}$ are shown as gray boxes outlining the corresponding rows and columns. From the covariance heatmaps (left), variation in MandibleLength(R) (the 14th variable) produces very little change in TO for females but substantial change for males. Among males, an increase in MandibleLength(R) is associated with marked increases in TO areas (the 7th and 8th variables). This association is faint when only the most related variables between the skull and TO are selected. Similarly, the 5th and 9th skull variables (PlToPr and RamusWidth(L)) display sex-specific patterns and are also selected by our CCR model with $(s_{1},s_{2})=(5,5)$.

For additional biomechanical interpretation, we use the joint reaction force (JRF), calculated as the residual force at the TMJ required to maintain static equilibrium under estimated muscle forces during mandibular motion \citep{she2021sexual}; the JRF magnitude is defined as the length of this vector. A larger JRF indicates greater loading on the TMJ, and for the same level of bite force, JRF increases as 3D mandibular length decreases \citep{sun2024explainable}. In raw-scale one-to-one comparisons, females with shorter mandibular length and smaller attachment areas exhibit larger JRF magnitudes, resulting in overall higher JRF values than males. Consistently, our result in \eqref{M-eq:linear_combination_TMJ} shows that females tend to have shorter mandibular length (MandibleLength (R)) as the selected size-related variables in TO increase, when other skull variables are held fixed, highlighting a potential high-risk subgroup for TMD that may not be evident in raw-scale pairwise analyses. These findings suggest that the CCR model offers new insights into the relationship between skull geometry and muscle attachment, pointing to future research directions on abnormal or high-risk subgroups and their biomechanical implications. Supporting results for marginally standardized data are provided in Section \ref{S-sec:skullvsTO_standardized}, and a raw-scale plot of mandibular length, muscle attachment area, and JRF is shown in Figure \ref{S-fig:plot_skull_vs_TO_areaL} of the Supplementary Materials.

 \section{DISCUSSION}\label{M-sec:discussion}
 
This paper proposes the conditional cross-covariance reduction model, which is easy to interpret and is designed to glean new information on the dynamic relationship of two sets of variables conditioning on the third binary variable. Instead of penalizing nonzero components, we apply hard-thresholding to them and introduce a sequential permutation for selecting sparsity method  under limited sample sizes, which is practical for the small sample size of the skull and muscle attachment measurements of the TMJ.

The conditional cross-covariance reduction model can be extended to a tensor, or multi-array, formulation to accommodate repeated measures and multiway structured data, thereby capturing both temporal and modality-specific dynamic associations. While the sequential permutation for selecting sparsity method provides a robust, data-driven strategy, it could be further enhanced to quantify uncertainty in the selected features—for instance, by incorporating permutation-based confidence intervals or resampling methods to assess variability. At present, the p-values from this procedure yield a single decision rather than reflecting the full uncertainty range of the estimated effects. Developing such extensions would further strengthen the robustness of the conditional cross-covariance reduction modeling framework for longitudinal and multi-modal dynamic association analysis.

%
%
%
%



\section*{FUNDING}
The data collection was supported by National Institute of Dental and Craniofacial Research (NIDCR), National Institutes of Health (NIH) under grant number R01DE021134, while the analysis was supported by NIDCR, NIH under grant number R03DE030509.

\section*{DATA AVAILABILITY}
The data that support the findings of this study are available from the \href{https://haiyao.people.clemson.edu/index.html}{Tissue Biomechanics Lab}. Restrictions apply to the availability of these data, which were used under license for this study. The implementation code can be found online \href{https://github.com/sparkqkr/CCR}{github.com/sparkqkr/CCR}.


\vspace*{-8pt}



\bibliographystyle{biom.bst} 
\bibliography{CCR.bib}


\appendix
\section*{Supplementary Materials}

\renewcommand{\thetable}{S\arabic{table}}
\setcounter{table}{0}

\renewcommand{\thefigure}{S\arabic{figure}}
\setcounter{figure}{0}

\renewcommand{\thealgorithm}{S\arabic{algorithm}}
\setcounter{algorithm}{0}

\renewcommand{\theequation}{S\arabic{equation}}
\setcounter{equation}{0}

\section{Additional Simulation Results}\label{S-sec:additional_simulation}
\subsection{Simulation in Different Group Correlation}
From a CCR model $\bolPhi$ in (4), we can control the group difference through $\rho_{1}$ and $\rho_{2}$ where $\rho_{1}-\rho_{2} >0$. If we have smaller values of $\rho_{1}-\rho_{2}$, we have small group differences, and it makes the CCR model hard to capture the group difference. 

In \ref{S-supptable:different_rhos}, we examined the CCR model in different values of $\rho_{1}$ and $\rho_{2}$. The worst case is $( \rho_{1}, \rho_{2})=( 0.25,-0.25 )$, $n_{1}=n_{2}=20$ which has the smallest group difference under the smallest sample size. In the worst result, the TPR values are large enough with smaller FPR values. And the subspace distances $D_{\mbV}$ and $D_{\mbU}$ are admittedly estimated ($<0.5$).We can compare the result of the associated correlation difference $\widehat{\eta}_{1}=1.12$ when $( \rho_{1}, \rho_{2})=( 0.6,-0.6 )$, $n_{1}=n_{2}=20$ to the $\widehat{\eta}_{1}=1.16$ in the real data analysis. The  $\widehat{\eta}$ values may support our real data analysis result by correctly selecting the variables on the skull and the temporalis origin (TO) muscle to specify the sex dimorphism in TMJ mechanics.

\begin{table}[H]
\centering
\caption{Numerical evaluations under rank-1 of cross-covariances in different $(\rho_{1},\rho_{2})$ over 100 data replicates. The numbers in parentheses report the standard error of the subspace distances $\text{D}_{\mbU}$, $\text{D}_{\mbV}$, covariance difference $\widehat{\delta}_{1}$, and the associated correlation differences $\widehat{\eta}_{1}$. Here, we used the true sparsity levels ($s_{1}^{*}=s_{1}$, $s_{2}^{*}=s_{2}$) for estimation.}
\label{S-supptable:different_rhos}
\begin{tabular}{ccccccc}
\multirow{2}{*}{\begin{tabular}[c]{@{}c@{}}Rank\\ ($r=1$)\end{tabular}} & \multicolumn{6}{c}{$(\rho_{1},\rho_{2})$}\\
& \multicolumn{2}{c}{$(0.25,-0.25)$} & \multicolumn{2}{c}{$(0.6,-0.6)$}     & \multicolumn{2}{c}{$(0.9,-0.9)$} \\ \hline \hline 
$n_{1}=n_{2}$                                                         & 20 & 200 & 20 & 200 & 20 & 200 \\ \hline
\rowcolor{Gray} $\text{TPR}_{\mbX}$                                                   & 0.777 & 0.997 & 0.983 & 1.000 & 1.000 & 1.000 \\
$\text{TPR}_{\mbY}$                                                   & 0.810 & 0.990 & 0.980 & 1.000 & 1.000 & 1.000 \\
\rowcolor{Gray} $\text{FPR}_{\mbX}$                                                   & 0.045 & 0.001 & 0.003 & 0.000 & 0.000 & 0.000 \\
$\text{FPR}_{\mbY}$                                                   & 0.048 & 0.002 & 0.005 & 0.000 & 0.000 & 0.000 \\
\rowcolor{Gray} $D_{\mbU}$                                                            & \begin{tabular}[c]{@{}c@{}}0.440\\ (0.034)\end{tabular} & \begin{tabular}[c]{@{}c@{}}0.098\\ (0.008)\end{tabular} & \begin{tabular}[c]{@{}c@{}}0.177\\ (0.014)\end{tabular} & \begin{tabular}[c]{@{}c@{}}0.043\\ (0.002)\end{tabular} & \begin{tabular}[c]{@{}c@{}}0.091\\ (0.005)\end{tabular}  & \begin{tabular}[c]{@{}c@{}}0.028\\ (0.001)\end{tabular}  \\
$D_{\mbV}$                                                              & \begin{tabular}[c]{@{}c@{}}0.409\\ (0.032)\end{tabular} & \begin{tabular}[c]{@{}c@{}}0.109\\ (0.011)\end{tabular} & \begin{tabular}[c]{@{}c@{}}0.177\\ (0.015)\end{tabular} & \begin{tabular}[c]{@{}c@{}}0.041\\ (0.002)\end{tabular} & \begin{tabular}[c]{@{}c@{}}0.094\\ (0.006)\end{tabular}  & \begin{tabular}[c]{@{}c@{}}0.028\\ (0.001)\end{tabular}  \\
\rowcolor{Gray} $\widehat{\delta}_{1}$                                                  & \begin{tabular}[c]{@{}c@{}}4.340\\ (0.144)\end{tabular} & \begin{tabular}[c]{@{}c@{}}3.426\\ (0.073)\end{tabular} & \begin{tabular}[c]{@{}c@{}}7.857\\ (0.223)\end{tabular} & \begin{tabular}[c]{@{}c@{}}8.051\\ (0.077)\end{tabular} & \begin{tabular}[c]{@{}c@{}}11.758\\ (0.252)\end{tabular} & \begin{tabular}[c]{@{}c@{}}12.099\\ (0.088)\end{tabular} \\
$\widehat{\eta}_{1}$                                                    & \begin{tabular}[c]{@{}c@{}}0.841\\ (0.014)\end{tabular} & \begin{tabular}[c]{@{}c@{}}0.507\\ (0.010)\end{tabular} & \begin{tabular}[c]{@{}c@{}}1.198\\ (0.018)\end{tabular} & \begin{tabular}[c]{@{}c@{}}1.191\\ (0.006)\end{tabular} & \begin{tabular}[c]{@{}c@{}}1.786\\ (0.006)\end{tabular}  & \begin{tabular}[c]{@{}c@{}}1.794\\ (0.002)\end{tabular} 
\end{tabular}
\end{table}

\subsection{Simulation in Different Signal Strength}
Another parameter we can control is the ratio of $c_{1}/c_{2}$, which we described in Section 3.1. The ratio controls the strength of signals. For example, if we use $c_{1}/c_{2} > 1$, the signals for the non-zero components in $\bolSigma_{\mbX,1}$ and $\bolSigma_{\mbY,1}$ are larger than the signals for the zero components in $\bolSigma_{\mbX,2}$ and $\bolSigma_{\mbY,2}$, and it makes easy to estimate the non-zero components in $\mbX$ and $\mbY$, respectively. 

Thus, we simulated four different ratios as $c_{1}/c_{2} \in \{0.5,1,3,5\}$ and the corresponding result are tabulated in Table \ref{S-supptable:diff_ratios}. When we have weak signals ($c_{1}/c_{2}=0.5$) with a small sample size ($n_{1}=n_{2}=20$), the CCR model still well captures the true non-zero components ($\text{TPR}_{\mbX},\text{TPR}_{\mbY}>0.74$) and the subspace distances are still well estimated ($D_{\mbV}, D_{\mbU} <0.4$). We can compare the maximal covariance difference $\widehat{\delta}_{1}$ to the $\widehat{\delta}_{1}=119.65$ in the real data analysis. That these values are much larger than those in Table \ref{S-supptable:diff_ratios} supports that the results in Section 4 successfully differentiate the group difference through the CCR model.

\begin{table}[H]
\centering
\caption{Numerical evaluations under rank-1 of cross-covariances in different $c_{1}/c_{2} \in \{0.5,1,3,5\}$ over 100 data replicates. The numbers in parentheses report the standard error of the subspace distances $\text{D}_{\mbU}$, $\text{D}_{\mbV}$, covariance difference $\widehat{\delta}_{1}$, and the associated correlation differences $\widehat{\eta}_{1}$. Here, we used the true sparsity levels ($s_{1}^{*}=s_{1}=3$, $s_{2}^{*}=s_{2}=3$) for estimation.}
\label{S-supptable:diff_ratios}
\begin{tabular}{ccccccccc}
\multirow{2}{*}{\begin{tabular}[c]{@{}c@{}}Rank\\ ($r=1$)\end{tabular}} & \multicolumn{8}{c}{$c_{1}/c_{2}$} \\
& \multicolumn{2}{c}{0.5} & \multicolumn{2}{c}{1} & \multicolumn{2}{c}{3} & \multicolumn{2}{c}{5} \\ \hline \hline 
$n_{1}=n_{2}$ & 20 & 200 & 20 & 200 & 20 & 200 & 20 & 200 \\ \hline 
\rowcolor{Gray} $\text{TPR}_{\mbX}$                                                   & 0.747& 1.000 & 1.000 & 1.000 & 1.000 & 1.000 & 1.000 & 1.000 \\
$\text{TPR}_{\mbY}$                                                   & 0.750 & 1.000 & 0.997 & 1.000 & 1.000 & 1.000 & 1.000 & 1.000 \\
\rowcolor{Gray} $\text{FPR}_{\mbX}$                                                   & 0.051 & 0.000 & 0.000 & 0.000 & 0.000 & 0.000 & 0.000 & 0.000 \\
$\text{FPR}_{\mbY}$                                                   & 0.062 & 0.000 & 0.001 & 0.000 & 0.000 & 0.000 & 0.000 & 0.000 \\
\rowcolor{Gray} $D_{\mbU}$                                                            & \begin{tabular}[c]{@{}c@{}}0.385\\ (0.040)\end{tabular} & \begin{tabular}[c]{@{}c@{}}0.027\\ (0.002)\end{tabular} & \begin{tabular}[c]{@{}c@{}}0.094\\ (0.005)\end{tabular} & \begin{tabular}[c]{@{}c@{}}0.027\\ (0.002)\end{tabular} & \begin{tabular}[c]{@{}c@{}}0.091\\ (0.005)\end{tabular}  & \begin{tabular}[c]{@{}c@{}}0.028\\ (0.001)\end{tabular}  & \begin{tabular}[c]{@{}c@{}}0.093\\ (0.005)\end{tabular}  & \begin{tabular}[c]{@{}c@{}}0.026\\ (0.001)\end{tabular}  \\
$D_{\mbV}$                                                              & \begin{tabular}[c]{@{}c@{}}0.375\\ (0.040)\end{tabular} & \begin{tabular}[c]{@{}c@{}}0.031\\ (0.002)\end{tabular} & \begin{tabular}[c]{@{}c@{}}0.094\\ (0.008)\end{tabular} & \begin{tabular}[c]{@{}c@{}}0.031\\ (0.002)\end{tabular} & \begin{tabular}[c]{@{}c@{}}0.094\\ (0.095)\end{tabular}  & \begin{tabular}[c]{@{}c@{}}0.028\\ (0.001)\end{tabular}  & \begin{tabular}[c]{@{}c@{}}0.095\\ (0.005)\end{tabular}  & \begin{tabular}[c]{@{}c@{}}0.028\\ (0.001)\end{tabular}  \\
\rowcolor{Gray} $\widehat{\delta}_{1}$                                                  & \begin{tabular}[c]{@{}c@{}}2.145\\ (0.047)\end{tabular} & \begin{tabular}[c]{@{}c@{}}2.017\\ (0.015)\end{tabular} & \begin{tabular}[c]{@{}c@{}}3.920\\ (0.084)\end{tabular} & \begin{tabular}[c]{@{}c@{}}4.033\\ (0.029)\end{tabular} & \begin{tabular}[c]{@{}c@{}}11.758\\ (0.252)\end{tabular} & \begin{tabular}[c]{@{}c@{}}12.099\\ (0.088)\end{tabular} & \begin{tabular}[c]{@{}c@{}}19.588\\ (0.419)\end{tabular} & \begin{tabular}[c]{@{}c@{}}20.179\\ (0.144)\end{tabular} \\
$\widehat{\eta}_{1}$                                                    & \begin{tabular}[c]{@{}c@{}}1.650\\ (0.026)\end{tabular} & \begin{tabular}[c]{@{}c@{}}1.794\\ (0.002)\end{tabular} & \begin{tabular}[c]{@{}c@{}}1.783\\ (0.007)\end{tabular} & \begin{tabular}[c]{@{}c@{}}1.794\\ (0.002)\end{tabular} & \begin{tabular}[c]{@{}c@{}}1.786\\ (0.006)\end{tabular}  & \begin{tabular}[c]{@{}c@{}}1.794\\ (0.002)\end{tabular}  & \begin{tabular}[c]{@{}c@{}}1.786\\ (0.006)\end{tabular}  & \begin{tabular}[c]{@{}c@{}}1.796\\ (0.002)\end{tabular} 
\end{tabular}
\end{table}

\subsection{Simulation in Different Covariance Structure}
Here, we added the simulation results in different covariance structures in data generation since data in Section 3 were generated from the autoregressive structure with parameter 0.7. 

We used two different covariance structures (compound symmetric, autoregressive) with parameters in $\{0.3,0.6,0.9\}$. The results are displayed in Table \ref{S-supptable:diff_cov_structure}. There is no substantial difference in the different covariance structures. We could see that the maximal covariance difference $\widehat{\delta}_{1}$ increases as the parameter increases since the larger parameter generates larger eigenvalues on the corresponding covariance matrices for data generation.

\begin{table}[H]
\centering
\caption{Numerical evaluations under rank-1 of cross-covariances in different covariance structures (auto-regressive and compound symmetric structures) with parameter $\{0.3, 0.6, 0.9\}$ over 100 data replicates. The numbers in parentheses report the standard error of the subspace distances $\text{D}_{\mbU}$, $\text{D}_{\mbV}$, covariance difference $\widehat{\delta}_{1}$, and the associated correlation differences $\widehat{\eta}_{1}$. Here, we used the true sparsity levels ($s_{1}^{*}=s_{1}$, $s_{2}^{*}=s_{2}$) for estimation.}
\label{S-supptable:diff_cov_structure}
\begin{tabular}{ccccccc}
\begin{tabular}[c]{@{}c@{}}Rank\\ ($r=1$)\end{tabular} & \multicolumn{3}{c}{compound symmetric (CS)} & \multicolumn{3}{c}{auto-regressive (AR)} \\ \hline \hline 
Parameter & 0.3 & 0.6 & 0.9 & 0.3 & 0.6 & 0.9 \\ \hline
\rowcolor{Gray} $\text{TPR}_{\mbX}$                                                   & 1.000 & 1.000 & 1.000 & 1.000 & 1.000 & 1.000 \\
$\text{TPR}_{\mbY}$                                                   & 1.000 & 1.000 & 1.000 & 1.000 & 1.000 & 1.000 \\
\rowcolor{Gray} $\text{FPR}_{\mbX}$                                                   & 0.000 & 0.000 & 0.000 & 0.000 & 0.000 & 0.000 \\
$\text{FPR}_{\mbY}$                                                   & 0.000 & 0.000 & 0.000 & 0.000 & 0.000 & 0.000 \\
\rowcolor{Gray} $D_{\mbU}$                                                            & \begin{tabular}[c]{@{}c@{}}0.039\\ (0.002)\end{tabular} & \begin{tabular}[c]{@{}c@{}}0.033\\ (0.002)\end{tabular} & \begin{tabular}[c]{@{}c@{}}0.028\\ (0.001)\end{tabular} & \begin{tabular}[c]{@{}c@{}}0.033\\ (0.002)\end{tabular} & \begin{tabular}[c]{@{}c@{}}0.027\\ (0.002)\end{tabular}  & \begin{tabular}[c]{@{}c@{}}0.023\\ (0.001)\end{tabular}  \\
$D_{\mbV}$                                                              & \begin{tabular}[c]{@{}c@{}}0.036\\ (0.002)\end{tabular} & \begin{tabular}[c]{@{}c@{}}0.033\\ (0.002)\end{tabular} & \begin{tabular}[c]{@{}c@{}}0.027\\ (0.001)\end{tabular} & \begin{tabular}[c]{@{}c@{}}0.037\\ (0.002)\end{tabular} & \begin{tabular}[c]{@{}c@{}}0.030\\ (0.002)\end{tabular}  & \begin{tabular}[c]{@{}c@{}}0.023\\ (0.001)\end{tabular}  \\
\rowcolor{Gray} $\widehat{\delta}_{1}$                                                  & \begin{tabular}[c]{@{}c@{}}17.243\\ (0.127)\end{tabular} & \begin{tabular}[c]{@{}c@{}}19.897\\ (0.143)\end{tabular} & \begin{tabular}[c]{@{}c@{}}22.559\\ (0.163)\end{tabular} & \begin{tabular}[c]{@{}c@{}}16.616\\ (0.124)\end{tabular} & \begin{tabular}[c]{@{}c@{}}19.189\\ (0.138)\end{tabular} & \begin{tabular}[c]{@{}c@{}}22.295\\ (0.160)\end{tabular} \\
$\widehat{\eta}_{1}$                                                    & \begin{tabular}[c]{@{}c@{}}1.798\\ (0.002)\end{tabular} & \begin{tabular}[c]{@{}c@{}}1.797\\ (0.002)\end{tabular} & \begin{tabular}[c]{@{}c@{}}1.797\\ (0.002)\end{tabular} & \begin{tabular}[c]{@{}c@{}}1.793\\ (0.002)\end{tabular} & \begin{tabular}[c]{@{}c@{}}1.793\\ (0.002)\end{tabular}  & \begin{tabular}[c]{@{}c@{}}1.796\\ (0.002)\end{tabular} 
\end{tabular}
\end{table}

\subsection{Simulation in Multi-Categorical Case}
We explore a case where the third variable $Z$ has more than two categories, multi-categorical variable. Unlike the binary case presented in the CCR model, we maximize each pairwise covariance difference by stacking the corresponding data matrices for $\mbX$ and $\mbY$, and estimate $\mbU$ and $\mbV$ separately for each pairwise comparison. 

For simplicity, we assume that $Z$ has three categories. Under the rank-1 scenario, we construct the pairwise covariance differences as follows: 
$$\bolPhi_{12}\bolPhi_{12}^{\top}=(\rho_{1}-\rho_{2})^{2}, \enspace \bolPhi_{23}\bolPhi_{23}^{\top}=(\rho_{2}-\rho_{3})^{2}, \enspace \bolPhi_{31}\bolPhi_{31}^{\top}=(\rho_{3}-\rho_{1})^{2},$$
where $\rho_{1}$, $\rho_{2}$, and $\rho_{3}$ represent group-specific canonical correlations, and each pairwise difference (e.g., $\rho_{1}-\rho_{2}$) corresponds to the singular value in the SVD of the associated pairwise difference matrix. Then, we construct the sample estimates by stacking the pairwise covariance difference matrices as follows:
\begin{align*}
    \widetilde{\bolPhi}_{\mbX}=[\widetilde{\bolPhi}_{12}, \widetilde{\bolPhi}_{23},\widetilde{\bolPhi}_{31}] \in \mathbb{R}^{p_{1} \times (3\cdot p_{2})}, \enspace \widetilde{\bolPhi}_{\mbY}=[\widetilde{\bolPhi}_{12}^{\top}, \widetilde{\bolPhi}_{23}^{\top},\widetilde{\bolPhi}_{31}^{\top}]^{\top} \in \mathbb{R}^{(3\cdot p_{1}) \times p_{2}},
\end{align*}
where $\widetilde{\bolPhi}_{12}$, $\widetilde{\bolPhi}_{23}$, and $\widetilde{\bolPhi}_{31}$ denote the pairwise sample covariance difference matrices. We then estimate the sparse matrices $\mbU$ and $\mbV$ following the procedure described in Algorithm \ref{S-alg:CCR_multigroup}.

\begin{algorithm}[H]
\caption{CCR model for multi-categorical $Z$ via two-way iterative thresholding}
\begin{algorithmic}[1]
\Inputs{The sample estimate $\widetilde{\bolPhi}_{\mbX} \in \mathbb{R}^{p_{1} \times (3\cdot p_{2})}$ and $\widetilde{\bolPhi}_{\mbY} \in \mathbb{R}^{(3 \cdot p_{1}) \times p_{2}}$, the corresponding rank $r \leq \mathrm{min}(p_{1},p_{2})$, and the sparsity levels $s_{1} \leq p_{1}$, $s_{2} \leq p_{2}$.}
\medskip
\Initialize{From $\widetilde{\bolPhi}$, calculate the initial top-left $r$ vectors of orthonormal matrix $\widehat{\mbV}^{(0)} =\text{SVD}\{ \widetilde{\bolPhi}_{\mbY}^{\top}\} \in \mathbb{R}^{p_2 \times r}$. Likewise, calculate the initial top-left $r$ vectors of orthonormal matrix $\widehat{\mbU}^{(0)} =\text{SVD}\{ \widetilde{\bolPhi}_{\mbX}\} \in \mathbb{R}^{p_1 \times r}$.
}
\State \textbf{Repeat $t=1,2,\ldots$}
\begin{enumerate}[label=(\alph*)]
\item Left multiplication: $\mbU^{(t),\text{mul}}=\widetilde{\bolPhi}_{\mbX}[\widehat{\mbV}^{(t-1)\top},\widehat{\mbV}^{(t-1)\top},\widehat{\mbV}^{(t-1)\top}]^{\top}.$
\item Left thresholding: for $I \subseteq \{1,2,\ldots,p_{1}\}$ and $i=1,\ldots,p_{1}$,
\begin{equation*}
    \mbU_{i}^{(t),\text{thr}}=
    \begin{cases}
        \mbU_{i}^{(t),\text{mul}} & , i \in \{\argmax_{|I|=s_{1}} \sum_{l \in I} \|\mbU_{l}^{(t),\text{mul}}\|_{2}\} \\
        0 & , \text{otherwise}
    \end{cases}
\end{equation*}

\item Left orthonormalization: QR decomposition on $\mbU^{(t),\text{thr}}$,
\begin{equation*}
\text{such that} \ \widehat{\mbU}^{(t)} \ \text{satisfies}\ \mathrm{span}(\widehat{\mbU}^{(t)})=\mathrm{span}(\widehat{\mbU}^{(t),\text{thr}}) \  \text{when} \ \{\widehat{\mbU}^{(t)}\}^{\top}\widehat{\mbU}^{(t)}=\mbI_{r}.     
\end{equation*}
\item Right multiplication:: $\mbV^{(t),\text{mul}}=\widetilde{\bolPhi}_{\mbY}^{\top}[\widehat{\mbU}^{(t)\top},\widehat{\mbU}^{(t)\top},\widehat{\mbU}^{(t)\top}]^{\top}.$
\item Right thresholding: for $J \subseteq \{1,2,\ldots,p_{2}\}$ and $j=1,\ldots,p_{2}$,
\begin{equation*}
    \mbV_{j}^{(t),\text{thr}}=
    \begin{cases}
        \mbV_{j}^{(t),\text{mul}} & , j \in \{\argmax_{|J|=s_{2}} \sum_{l \in J} \|\mbV_{l}^{(t),\text{mul}}\|_{2}\} \\
        0 & , \text{otherwise}
    \end{cases}
\end{equation*}

\item Right orthonormalization: QR decomposition on $\mbV^{(t),\text{thr}}$,
\begin{equation*}
\text{such that} \ \widehat{\mbV}^{(t)} \ \text{satisfies}\ \mathrm{span}(\widehat{\mbV}^{(t)})=\mathrm{span}(\widehat{\mbV}^{(t),\text{thr}}) \  \text{when} \ \{\widehat{\mbV}^{(t)}\}^{\top}\widehat{\mbV}^{(t)}=\mbI_{r}.     
\end{equation*}
\end{enumerate}
\textbf{until} convergence.
\Output{$\widehat{\mbU}=\widehat{\mbU}^{(t)}$, $\widehat{\mbV}=\widehat{\mbV}^{(t)}$.}

\end{algorithmic}
\label{S-alg:CCR_multigroup}
\end{algorithm}

We evaluate the estimation performance of the proposed method through numerical experiments. In the three-group example, the total sample size is $N=n_{1}+n_{2}+n_{3}$. For $i=1,\ldots,n_{1}$, we generate $(\mbx_{i},\mby_{i})$ jointly from a normal distribution with mean zero and covariance $\bolSigma_{1}$. For $i=n_{1}+1,\ldots,n_{1}+n_{2}$, we generate $(\mbx_{i},\mby_{i})$ jointly from a normal distribution with mean zero and covariance $\bolSigma_{2}$. Lastly, for $i=n_{1}+n_{2}+1,\ldots,N$, we generate $(\mbx_{i},\mby_{i})$ jointly from a normal distribution with mean zero and covariance $\bolSigma_{3}$ where the covariance matrix for each group is as follows:
\begin{align*}
    \bolSigma_{z}=\begin{pmatrix}
        \bolSigma_{\mbX} & \rho_{z} \mbU \mbV^{\top} \\
        \rho_{z}\mbV \mbU^{\top} & \bolSigma_{\mbY}
     \end{pmatrix}, \enspace z=1,2,3,
\end{align*}
with the group index $z$ represents the categorical variable $Z \in \{1,2,3 \}$. The $\mbU$ and $\mbV$ under rank-1 scenario are the same as in Section 3. We set $p_{1}=18$, $p_{2}=15$, $s_{1}=3$, $s_{2}=3$, $n_{1}=n_{2}=n_{3} \in \{20,200\}$, $c_{1}/c_{2}=3$ and the group difference $(\rho_{1},\rho_{2},\rho_{3}) \in \{(0.9,0.1,-0.9),(0.9,-0.4,-0.5)\}$. Table \ref{S-supptable:three_group_result} presents the results for the multi-category $Z$ case over 100 replicated datasets. The CCR model successfully identifies the nonzero variables in $\mbX$ and $\mbY$, as reflected in the high TPRs and low FPRs. Since we maximize each pairwise covariance difference, we report the estimated correlations for each linear combination, defined as $\rho_{k}=\Corr (\mbU\mbX_{k},\mbV\mbY_{k})$, $k=1,2,3$, instead of directly presenting $\widehat{\delta}_{1}$ and $\widehat{\eta}_{1}$.
The accurate estimation of $\widehat{\rho}_{1}$, $\widehat{\rho}_{2}$, and $\widehat{\rho}_{3}$  indicates that the group-specific differences are effectively captured, reflecting strong performance in estimating $\widehat{\eta}_{1}$.  

\begin{table}[H]
\centering
\caption{Numerical evaluations under rank-1 of cross-covariances in different $(\rho_{1},\rho_{2},\rho_{3} )$ over 100 data replicates. The numbers in parentheses report the standard error of the subspace distances $\text{D}_{\mbU}$, $\text{D}_{\mbV}$, covariance difference $\widehat{\delta}_{1}$, and the associated correlation differences $\widehat{\eta}_{1}$. Here, we used the true sparsity levels ($s_{1}^{*}=s_{1}$, $s_{2}^{*}=s_{2}$) for estimation.}
\label{S-supptable:three_group_result}
\begin{tabular}{ccccc}
\multirow{2}{*}{\begin{tabular}[c]{@{}c@{}}Rank\\ ($r=1$)\end{tabular}} & \multicolumn{4}{c}{$(\rho_{1},\rho_{2},\rho_{3})$}\\
& \multicolumn{2}{c}{$(0.9,0.1,-0.9)$} & \multicolumn{2}{c}{$(0.9,-0.4,-0.5)$}    \\ \hline \hline 
$n_{1}=n_{2}$                                                         & 20 & 200 & 20 & 200 \\ \hline
\rowcolor{Gray} $\text{TPR}_{\mbX}$                                                   & 0.773 & 1.000 & 0.937 & 1.000 \\
$\text{TPR}_{\mbY}$                                                                   & 0.717 & 1.000 & 0.907 & 1.000 \\
\rowcolor{Gray} $\text{FPR}_{\mbX}$                                                   & 0.045 & 0.000 & 0.013 & 0.000 \\
$\text{FPR}_{\mbY}$                                                                   & 0.071 & 0.000 & 0.023 & 0.000 \\
\rowcolor{Gray} $D_{\mbU}$                                                            & \begin{tabular}[c]{@{}c@{}}0.447\\ (0.034)\end{tabular} & \begin{tabular}[c]{@{}c@{}}0.075\\ (0.004)\end{tabular} & \begin{tabular}[c]{@{}c@{}}0.216\\ (0.024)\end{tabular} & \begin{tabular}[c]{@{}c@{}}0.043\\ (0.002)\end{tabular} \\
$D_{\mbV}$                                                              & \begin{tabular}[c]{@{}c@{}}0.486\\ (0.036)\end{tabular} & \begin{tabular}[c]{@{}c@{}}0.075\\ (0.004)\end{tabular} & \begin{tabular}[c]{@{}c@{}}0.256\\ (0.027)\end{tabular} & \begin{tabular}[c]{@{}c@{}}0.046\\ (0.003)\end{tabular}  \\
\rowcolor{Gray} $\widehat{\rho}_{1}$                                                  & \begin{tabular}[c]{@{}c@{}}0.717\\ (0.031)\end{tabular} & \begin{tabular}[c]{@{}c@{}}0.896\\ (0.001)\end{tabular} & \begin{tabular}[c]{@{}c@{}}0.830\\ (0.021)\end{tabular} & \begin{tabular}[c]{@{}c@{}}0.896\\ (0.001)\end{tabular} \\
$\widehat{\rho}_{2}$                                                  & \begin{tabular}[c]{@{}c@{}}-0.116\\ (0.021)\end{tabular} & \begin{tabular}[c]{@{}c@{}}0.098\\ (0.008)\end{tabular} & \begin{tabular}[c]{@{}c@{}}-0.396\\ (0.019)\end{tabular} & \begin{tabular}[c]{@{}c@{}}-0.403\\ (0.007)\end{tabular} \\
\rowcolor{Gray} $\widehat{\rho}_{3}$                                                  & \begin{tabular}[c]{@{}c@{}}-0.638\\ (0.042)\end{tabular} & \begin{tabular}[c]{@{}c@{}}-0.896\\ (0.001)\end{tabular} & \begin{tabular}[c]{@{}c@{}}-0.445\\ (0.022)\end{tabular} & \begin{tabular}[c]{@{}c@{}}-0.498\\ (0.005)\end{tabular}  
\end{tabular}
\end{table}

\begin{figure}[H]
     \centering
     \begin{subfigure}[b]{0.45\textwidth}
         \centering
         \includegraphics[width=\textwidth]{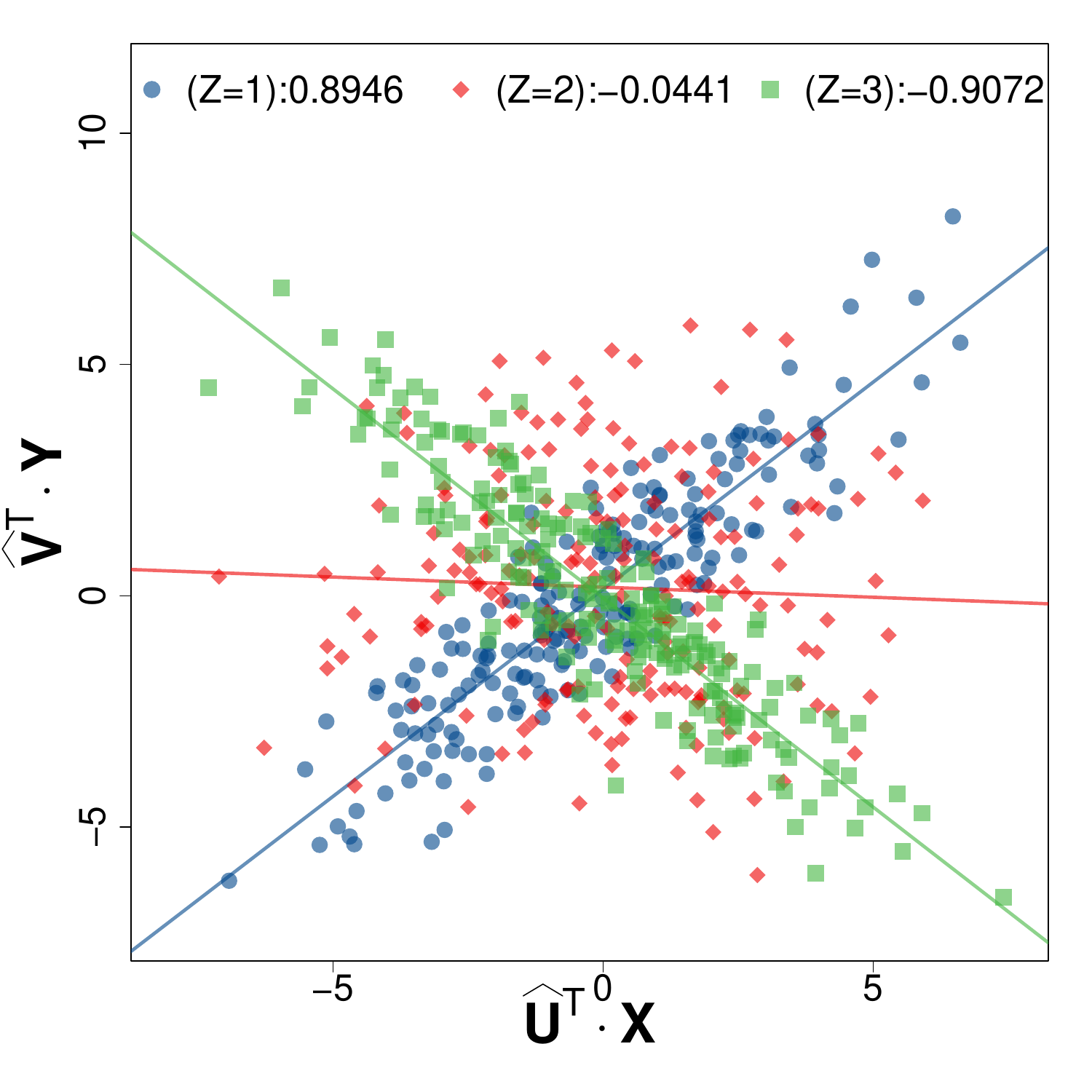}
         \caption{$(\rho_{1},\rho_{2},\rho_{3})=(0.9,0.1,-0.9)$}
     \end{subfigure}
     \begin{subfigure}[b]{0.45\textwidth}
         \centering
         \includegraphics[width=\textwidth]{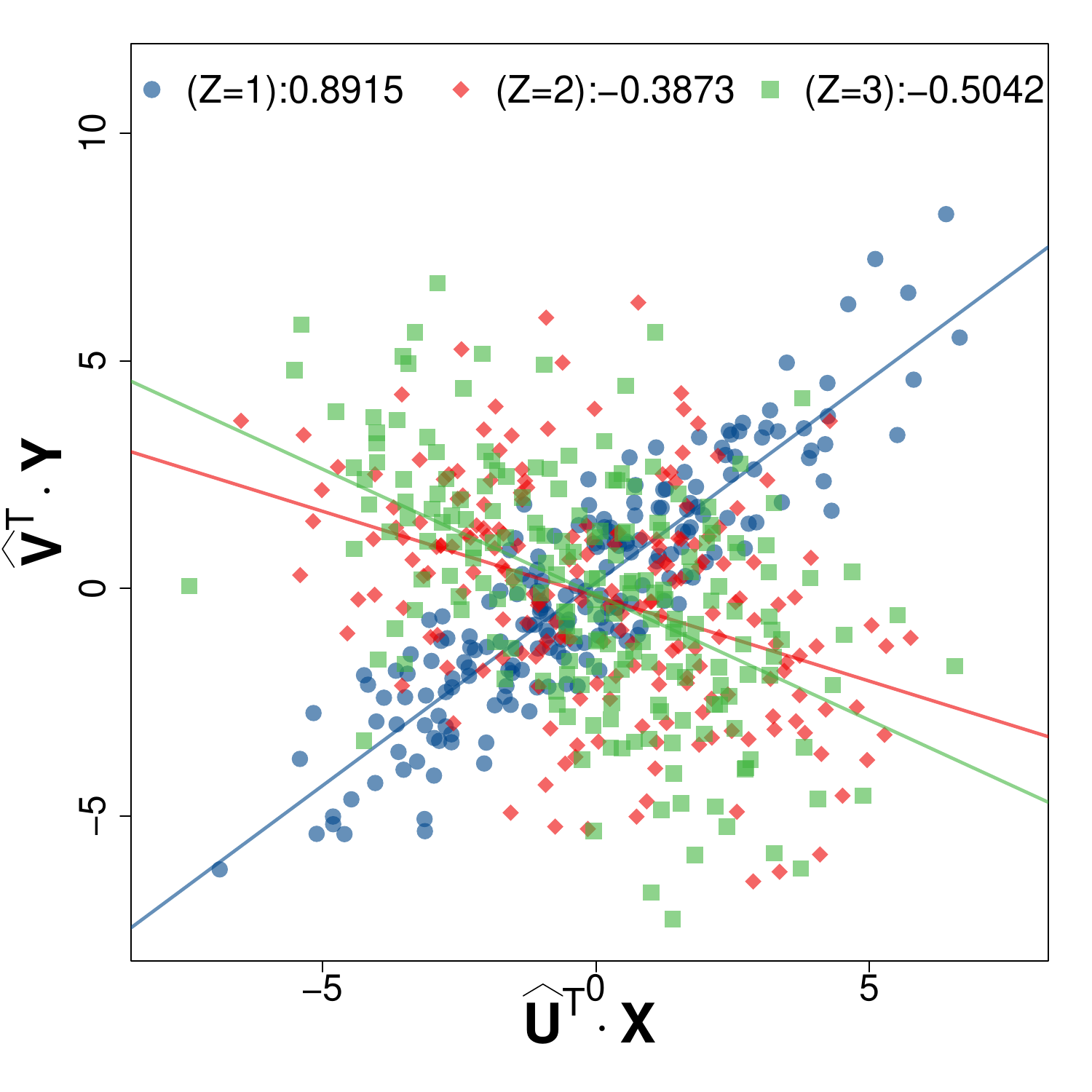}
         \caption{$(\rho_{1},\rho_{2},\rho_{3})=(0.9,-0.4,-0.5)$}
     \end{subfigure}
        \caption{Linear combinations of the CCR model for multi-group case in different group-specific correlations $(\rho_{1},\rho_{2},\rho_{3}) \in \{(0.9,0.1,-0.9),(0.9,-0.4,-0.5)\}$. The numbers in the legend represent the correlation between linear combinations by $Z \in \{1,2,3\}$. Sparsity levels are set as $\widehat{s}_{1}=\widehat{s}_{2}=3$ and $n_{1}=n_{2}=n_{3}=200$.}
        \label{S-fig:three_group_result}
\end{figure}

We present the linear combinations of $\mbX$ and $\mbY$ for the multi-group case in Figure \ref{S-fig:three_group_result}. The plot shows a clear separation among the groups—blue circles, red diamonds, and green squares—indicating that the CCR model effectively extends to multi-group settings by maximizing each pairwise difference.

\section{Comparison to Competing Method}\label{S-sec:competing_method}
We compared the CCR model with competing methods to illustrate how the dynamic association between $\mbX$ and $\mbY$ is affected by the binary variable $Z \in \{1,2\}$. For competing methods, we used a generalized liquid association analysis(GLAA; \citealt{li2023generalized}), Bayesian canonical correlation analysis (BCCA, \citealt{klami2013bayesian}), and a regularized generalized canonical correlation analysis (RGCCA; \citealt{tenenhaus2011regularized}). 

Similar to the CCR model, GLAA proposed to specify the dynamic association between $\mbX$ and $\mbY$ by differentiating the continuous $\mbZ \in \mathbb{R}^{p_{3}}$. With continuous $\mbZ$, GLAA can specify the dynamic association between $\mbX$ and $\mbY$ changes as $\mbZ$ varies. Thus, the GLAA result depends on the smooth differentiation with respect to $\mbZ$. BCCA extracts the statistical dependencies (correlations) between $\mbX$ and $\mbY$ but also decomposes $\mbX$ and $\mbY$ into shared and data-specific components. By introducing a latent variable that uses group-wise sparsity priors (spike-and-slab), BCCA specifies shared latent structure $\mbC$ and views specified latent variables $\mbC_{\mbX}$ and $\mbC_{\mbY}$. And those latent variables allow flexibility in capturing association patterns that vary across different groups. The latent modeling, so-called inter-battery factor analysis (IBFA; \citealt{browne1979maximum}), is denoted as $\mbX \sim \calN(\mbA_{\mbX}\mbC+\mbB_{\mbX}\mbC_{\mbX},\bolSigma_{\mbX})$ and $\mbY \sim \calN(\mbA_{\mbY}\mbC+\mbB_{\mbY}\mbC_{\mbY},\bolSigma_{\mbY})$ where $\mbA_{\mbX}$,$\mbA_{\mbY}$ are loading coefficients for the shared structure $\mbC$ and $\mbB_{\mbX}$,$\mbB_{\mbY}$ are loading coefficients for the individual structure $\mbC_{\mbX}$ and $\mbC_{\mbY}$, respectively. Lastly, RGCCA proposed to conduct multiblock analyses. For two datasets $\mbX$ and $\mbY$, RGCCA can be applied as sparse canonical correlation analysis, which maximizes the canonical correlation between $\mbX$ and $\mbY$ with sparsity on the canonical vectors. 

Similar to the simulation setting in Section 3. We set $p_{1}=18$, $p_{2}=15$, $s_{1}^{*}=3$, $s_{2}^{*}=3$, $N=n_{1}+n_{2}=20+20=40$. For data generation, $\rho_{1}=0.9$, $\rho_{2}=-0.9$, $c_{1}=3$, and  $c_{1}=1$ for data generation. We treated the binary variable as continuous to apply the GLAA and introduced sparsity only on $\mbX$ and $\mbY$. For BCCA, we set the number of canonical vectors to 1 and others from the default setting in \textit{R package ``CCAGFA''}. For RGCCA, we tuned the sparsity parameter in the range between $\min \{p_{1},p_{2}\}$ and one and used 0.33 as the sparsity parameter.

\begin{figure}[H]
     \centering
     \begin{subfigure}[b]{0.4\textwidth}
         \centering
         \includegraphics[width=\textwidth]{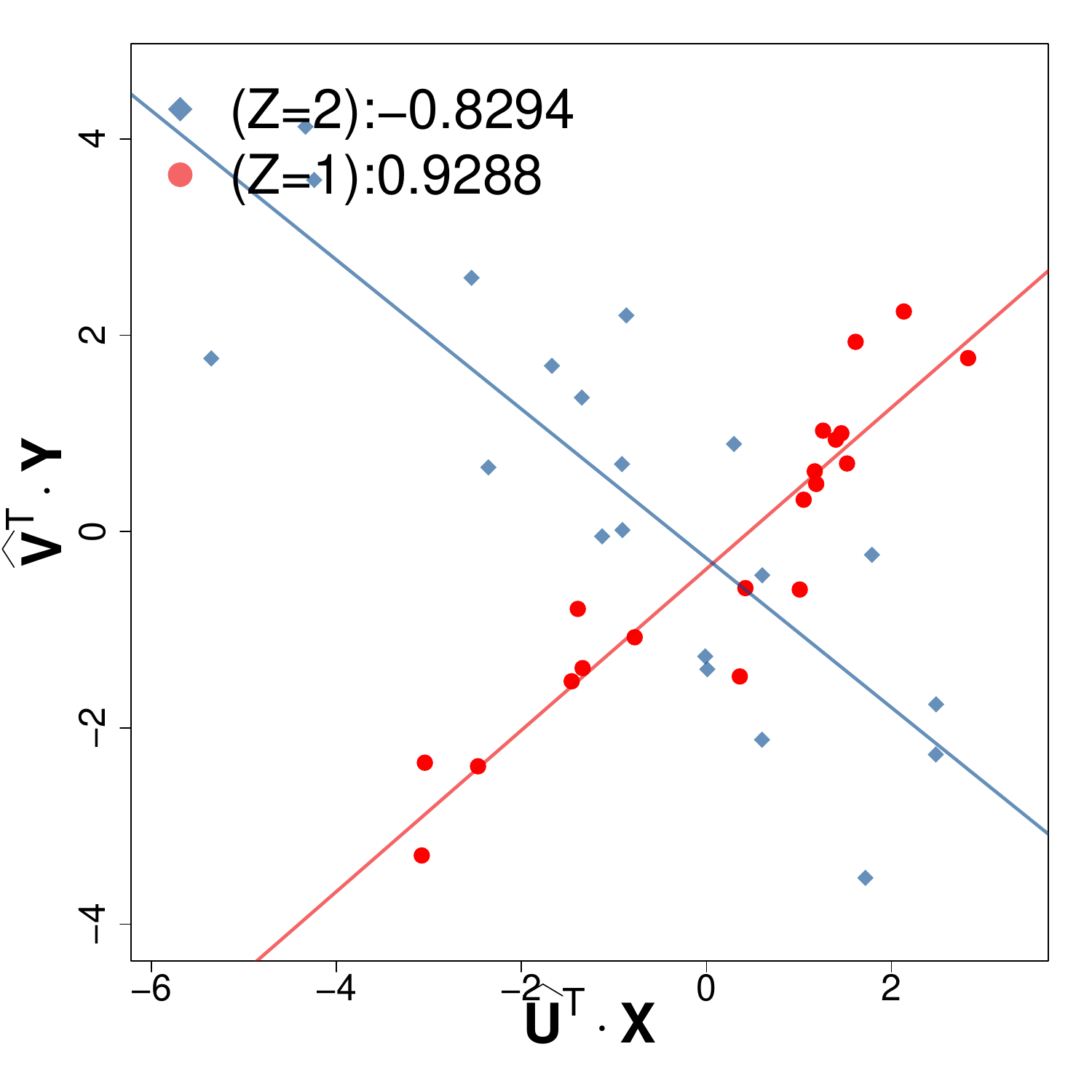}
         \caption{CCR}
        \end{subfigure}
        \begin{subfigure}[b]{0.4\textwidth}
         \centering
         \includegraphics[width=\textwidth]{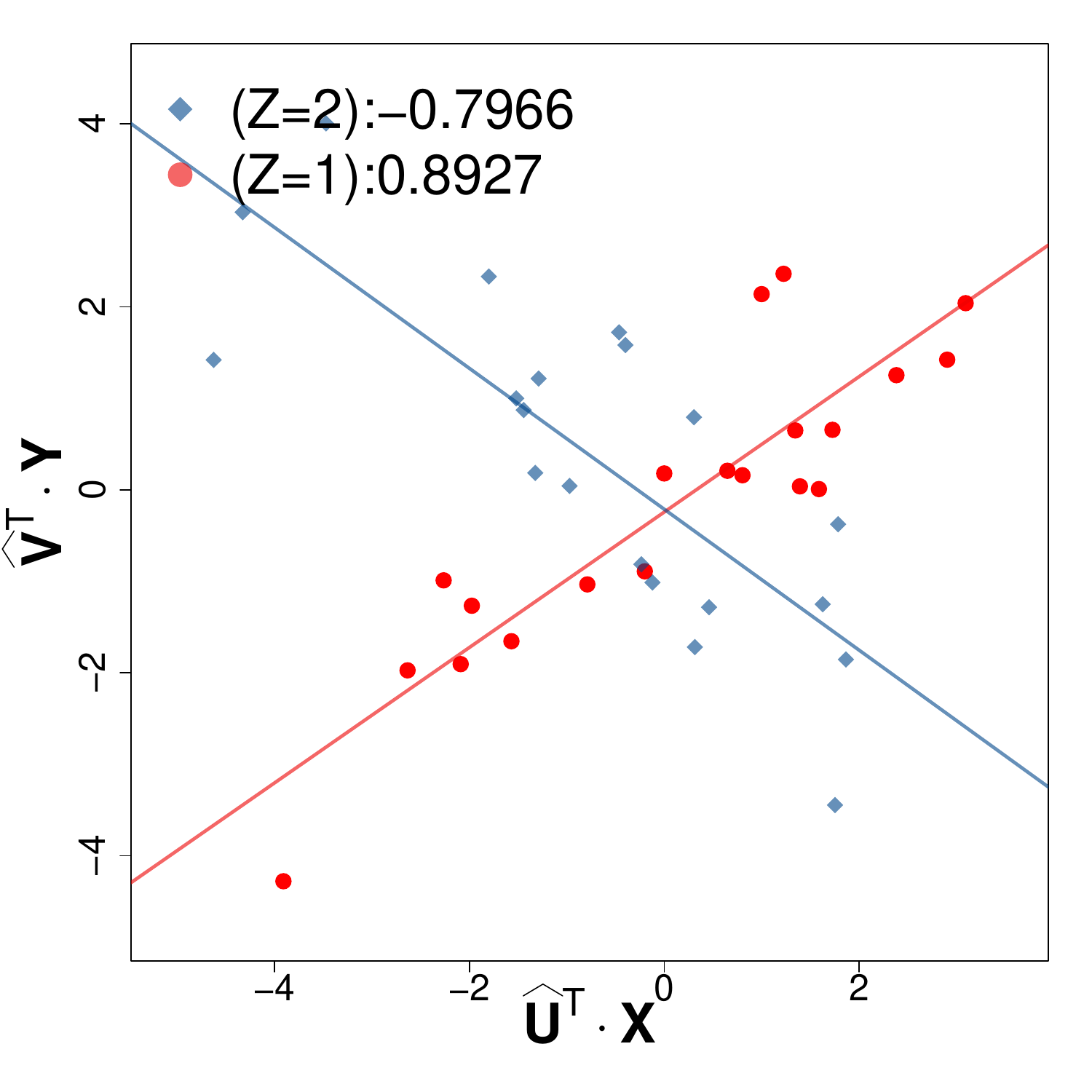}
         \caption{GLAA}
        \end{subfigure}
     
     \begin{subfigure}[b]{0.4\textwidth}
         \centering
         \includegraphics[width=\textwidth]{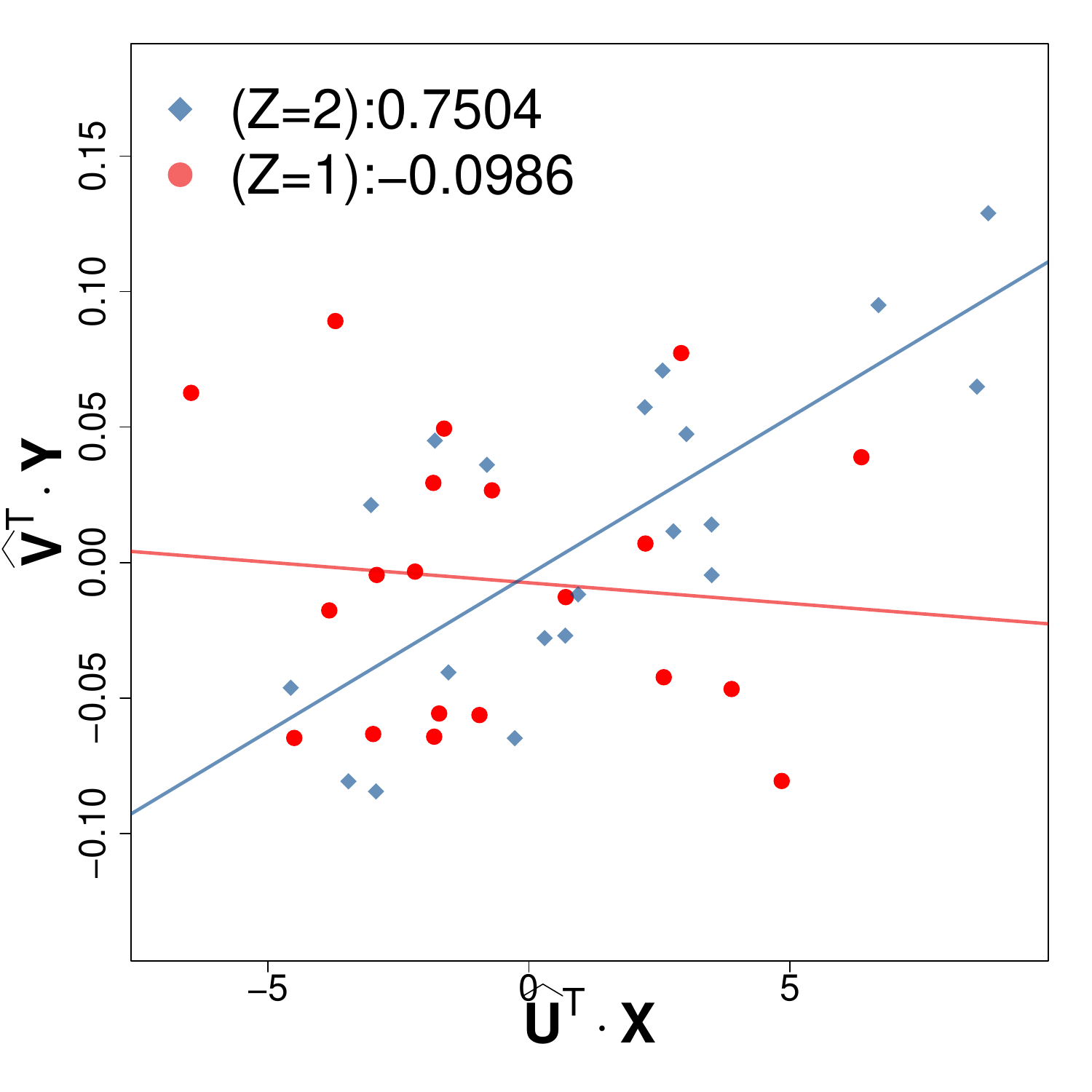}
         \caption{BCCA}
        \end{subfigure}
     \begin{subfigure}[b]{0.4\textwidth}
         \centering
         \includegraphics[width=\textwidth]{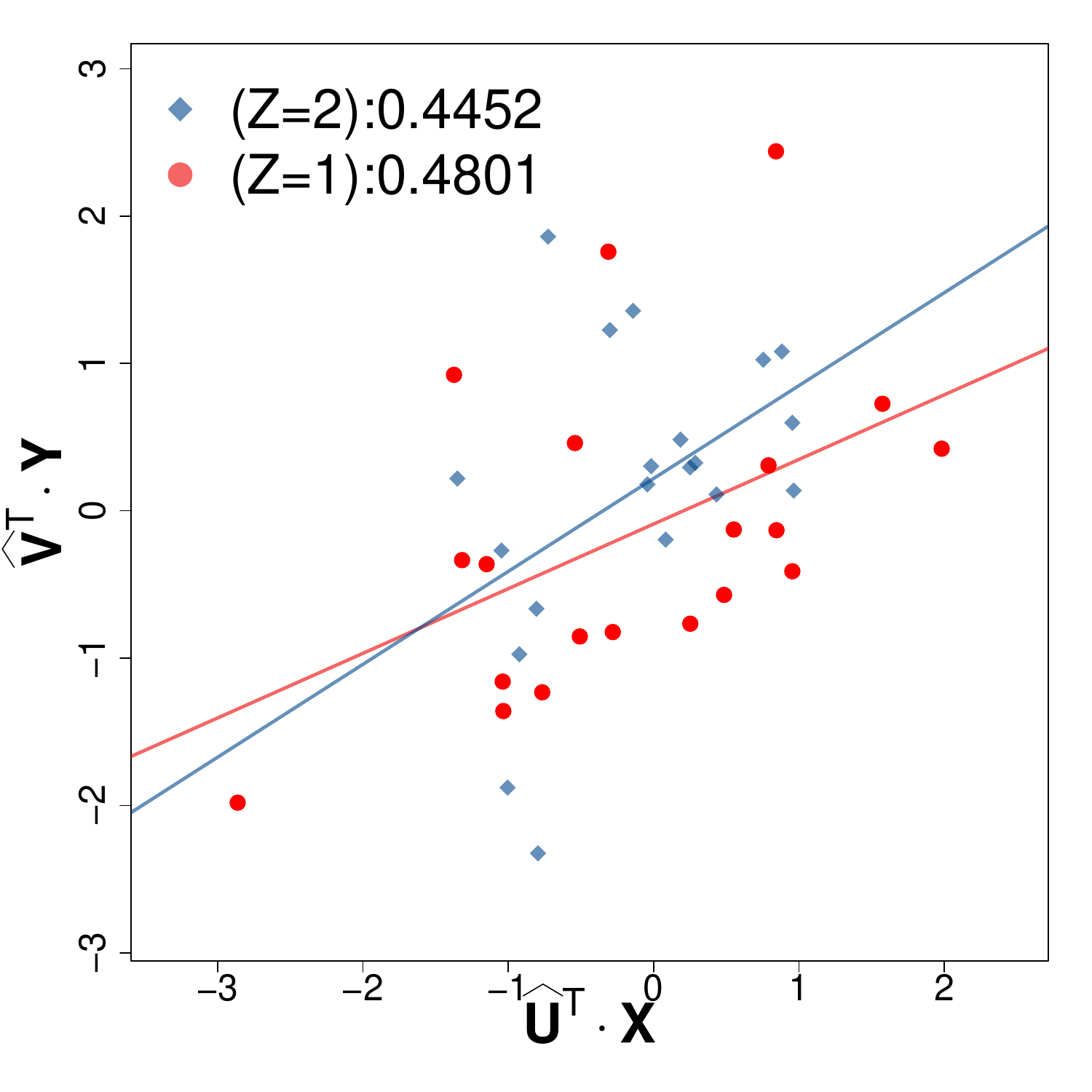}
         \caption{RGCCA}
        \end{subfigure}
        \caption{Comparison of dynamic association by binary variable $Z \in \{1,2\}$ with competing methods. The methods under comparison are: conditional cross-covariance reduction (CCR) model, generalized liquid association analysis (GLAA), Bayesian canonical correlation analysis (BCCA), and regularized generalized canonical correlation analysis (RGCCA).}
        \label{S-fig:comparison_competing}
\end{figure}

In Figure \ref{S-fig:comparison_competing}, we plotted the linear combinations of four methods on $\mbX$ and $\mbY$ and marked observations in different colors by group $Z$. The first three methods (CCR, GLAA, BCCA) extracted the dynamic association by $Z$. However, RGCCA cannot distinguish the difference between the group $Z$. The CCR model specifies the largest correlation difference of 1.76 among competing methods. The GLAA has the same estimation metric proportional to $\bolPhi$ in the CCR model. Thus, there is no significant difference in correlation between the CCR model and GLAA. However, since the GLAA is a penalized method, the result can be biased when estimating the canonical loadings. The CCR model reduces the bias through hard thresholding in the algorithm and estimates more accurate canonical loading vectors. The subspace distances $D_{\mbU}$ and $D_{\mbV}$ result in Table \ref{S-supptable:competing_methods} also support the less biased result in the CCR model.

We replicated the data generations 100 times and compared the results of the competing methods in Table \ref{S-supptable:competing_methods}. We used estimated loading vectors as $\widehat{\mbU}$ and $\widehat{\mbV}$ to calculate the subspace distances $D_{\mbU}$ and $D_{\mbV}$. First, among the competing methods, the CCR model has the best result on the variable selection. The CCR model and GLAA result select correct variables on $\mbX$ and $\mbY$ (based on $\text{TPR}_{\mbX}$, $\text{TPR}_{\mbY}$, $\text{FPR}_{\mbX}$, and $\text{FPR}_{\mbY}$). In BCCA, \cite{klami2013bayesian} noted that the elements of inactive components in the loading vectors are not forced exactly to zero but instead shrink toward minimal values under group-wise sparsity assumptions on the loading factors. They also introduced a group-wise spike-and-slab prior for stronger sparsity, but the corresponding implementation is not available. Therefore, we leave the TPRs and FPRs for BCCA blank in Table \ref{S-supptable:competing_methods}. Similar to Figure \ref{S-fig:comparison_competing}, the three methods (CCR, GLAA, BCCA) distinguish the group difference ($\widehat{\eta}_{1}$), with the CCR model achieving the smallest subspace distances.

\begin{table}[!ht]
\centering
\caption{Numerical evaluations of competing methods under rank-1 of cross-covariances with $c_{1}/c_{2}=3$ and $(\rho_{1},\rho_{2})=(0.9,-0.9)$ over 100 data replicates. The numbers in parentheses report the standard error of the subspace distances $\text{D}_{\mbU}$, $\text{D}_{\mbV}$, covariance difference $\widehat{\delta}_{1}$, and the associated correlation differences $\widehat{\eta}_{1}$. Here, we used the true sparsity levels ($s_{1}^{*}=s_{1}=3$, $s_{2}^{*}=s_{2}=3$) for estimation.}
\label{S-supptable:competing_methods}
\resizebox{\textwidth}{!}{
\begin{tabular}{cccccccccc}
Method & $n_{1},n_{2}$ & $\text{TPR}_{\mbX}$ & $\text{TPR}_{\mbY}$ & $\text{FPR}_{\mbX}$ & $\text{FPR}_{\mbY}$ & $D_{\mbU}$                & $D_{\mbV}$ & $\widehat{\delta}_{1}$ & $\widehat{\eta}_{1}$ \\ \hline \hline 
\multirow{3}{*}{CCR} & \cellcolor{Gray}20 & \cellcolor{Gray}1.000 & \cellcolor{Gray}1.000 & \cellcolor{Gray}0.000 & \cellcolor{Gray}0.000               & \cellcolor{Gray}\begin{tabular}[c]{@{}c@{}}0.093\\ (0.005)\end{tabular} &\cellcolor{Gray} \begin{tabular}[c]{@{}c@{}}0.098\\ (0.005)\end{tabular} &\cellcolor{Gray} \begin{tabular}[c]{@{}c@{}}12.597\\ (0.280)\end{tabular} & \cellcolor{Gray}\begin{tabular}[c]{@{}c@{}}1.798\\ (0.006)\end{tabular} \\
& 200 & 1.000 & 1.000 & 0.000 & 0.000 & \begin{tabular}[c]{@{}c@{}}0.040\\ (0.003)\end{tabular} & \begin{tabular}[c]{@{}c@{}}0.041\\ (0.002)\end{tabular} & \begin{tabular}[c]{@{}c@{}}12.168\\ (0.135)\end{tabular} & \begin{tabular}[c]{@{}c@{}}1.794\\ (0.003)\end{tabular} \\ \hline
 \multirow{3}{*}{GLAA}  & \cellcolor{Gray}20 & \cellcolor{Gray}1.000 &\cellcolor{Gray} 1.000 &\cellcolor{Gray} 0.269 &\cellcolor{Gray} 0.293               &\cellcolor{Gray} \begin{tabular}[c]{@{}c@{}}0.320\\ (0.009)\end{tabular} &\cellcolor{Gray} \begin{tabular}[c]{@{}c@{}}0.299\\ (0.010)\end{tabular} &\cellcolor{Gray} \begin{tabular}[c]{@{}c@{}}12.490\\ (0.283)\end{tabular} &\cellcolor{Gray} \begin{tabular}[c]{@{}c@{}}1.796\\ (0.006)\end{tabular} \\
 & 200 & 1.000 & 1.000 & 0.265 & 0.270 & \begin{tabular}[c]{@{}c@{}}0.152\\ (0.004)\end{tabular} & \begin{tabular}[c]{@{}c@{}}0.135\\ (0.004)\end{tabular} & \begin{tabular}[c]{@{}c@{}}12.146\\ (0.134)\end{tabular} & \begin{tabular}[c]{@{}c@{}}1.795\\ (0.003)\end{tabular} \\ \hline
\multirow{3}{*}{BCCA} &\cellcolor{Gray} 20 &\cellcolor{Gray} - &\cellcolor{Gray} - &\cellcolor{Gray} - &\cellcolor{Gray}  -               &\cellcolor{Gray} \begin{tabular}[c]{@{}c@{}}0.496\\ (0.028)\end{tabular} &\cellcolor{Gray} \begin{tabular}[c]{@{}c@{}}0.524\\ (0.031)\end{tabular} &\cellcolor{Gray} \begin{tabular}[c]{@{}c@{}}0.579\\ (2.283)\end{tabular}  &\cellcolor{Gray} \begin{tabular}[c]{@{}c@{}}1.264\\ (0.050)\end{tabular} \\
& 200 & - &  - & - & - & \begin{tabular}[c]{@{}c@{}}0.391\\ (0.034)\end{tabular} & \begin{tabular}[c]{@{}c@{}}0.427\\ (0.035)\end{tabular} & \begin{tabular}[c]{@{}c@{}}1.484\\ (1.516)\end{tabular}  & \begin{tabular}[c]{@{}c@{}}1.393\\ (0.045)\end{tabular} \\ \hline
\multirow{3}{*}{RGCCA} &\cellcolor{Gray} 20 &\cellcolor{Gray} 0.240 &\cellcolor{Gray} 0.527 &\cellcolor{Gray} 0.171 &\cellcolor{Gray} 0.072               &\cellcolor{Gray} \begin{tabular}[c]{@{}c@{}}0.931\\ (0.014)\end{tabular} &\cellcolor{Gray} \begin{tabular}[c]{@{}c@{}}0.797\\ (0.014)\end{tabular} &\cellcolor{Gray} \begin{tabular}[c]{@{}c@{}}0.574\\ (0.246)\end{tabular}  &\cellcolor{Gray} \begin{tabular}[c]{@{}c@{}}0.412\\ (0.043)\end{tabular} \\
& 200 & 0.290 & 0.597 & 0.156 & 0.057 & \begin{tabular}[c]{@{}c@{}}0.919\\ (0.014)\end{tabular} & \begin{tabular}[c]{@{}c@{}}0.769\\ (0.013)\end{tabular} & \begin{tabular}[c]{@{}c@{}}0.782\\ (0.269)\end{tabular}  & \begin{tabular}[c]{@{}c@{}}0.537\\ (0.054)\end{tabular}
\end{tabular}
}
\end{table}

 \section{Simulation on the SPSS Method}\label{S-sec:simulation}
We examine the accuracy of the SPSS method for selection of nonzero variables in $\mbX$ and $\mbY$. The simulation setup is the same as in Section 3.1.

 From the LTO data splits with $(n_{1},n_{2})=(10,11)$, we randomly flip the signs of the differences $D_{\ell}$, $\ell=1,2,\ldots,n_{1}n_{2}$ in each permutation under the null hypothesis $\text{H}_{0}: \overline{\delta}_{1}^{(i,k)} - \overline{\delta}_{1}^{(i+1,k)}=0$. We permute 1000 times to construct a reference distribution and compute the p-value.

 We replicate the data generation 100 times and evaluated the accuracy of the selected sparsity levels under varying signal strengths, $c_{1}/c_{2} \in \{3,5,7,10\}$. We tabulate the accuracies in Table \ref{S-table:accuracyofsparsity}. The accuracies tell us that the SPSS procedure guarantees that we select the true sparsity level when we have recognizable signals ($c_{1}/c_{2} \geq 5$). We apply SPSS procedure with in our TMJ application, as the actual differences in the TMJ data are greater than the simulated differences under $c_{1}/c_{2} = 10$.

\begin{table}[H]
\centering
\caption{Accuracies of the sparsity levels $s_{1}$, $s_{2}$ from the permutation test at different signal strengths $c_{1}/c_{2} \in \{3,5,7,10\}$.}
\normalsize
\begin{tabular}{ccc}
$c_{1}/c_{2}$ & Sparsity level & Accuracy \\ \hline \hline
\multirow{2}{*}{3} & $s_{1}$     & 0.69     \\ 
& $s_{2}$     & 0.63     \\  \hline
\multirow{2}{*}{5} & $s_{1}$     & 1      \\ 
& $s_{2}$     & 0.88       \\  \hline
\multirow{2}{*}{7} & $s_{1}$     & 1      \\ 
& $s_{2}$     & 1       \\  \hline
\multirow{2}{*}{10} & $s_{1}$     & 1      \\ 
& $s_{2}$     & 1       \\     
\end{tabular}
\label{S-table:accuracyofsparsity}
\end{table}

\section{An Alternative Criterion for Variable Selection}\label{S-sec:CCR_IC}
For the algorithm of the CCR model, we need to input the sparsity levels $s_{1}$ and $s_{2}$. Here, we construct a BIC-type criterion to estimate the sparsity levels. Similar to AIC and BIC, we use minus maximization criterion modified by a penalty derived from the model size. Our proposed information criterion is
$$\text{IC}\big\{s_{1},s_{2}\big\}=-N\cdot \log \big\{\|\mbP_{\widehat{\mbU}}\widetilde{\bolPhi}\mbP_{\widehat{\mbV}}\|_{\mathrm{F}}\big\}+(s_{1}+s_{2})\log(N).$$
The $\widetilde{\bolPhi}$ is the mean the sample covariance difference, $(s_{1}+s_{2})$ is the number of free parameters, $N=n_{1}+n_{2}$ is the number of data point, and $\widehat{\mbU}$ and $\widehat{\mbV}$ are estimated subspaces of the CCR model.

\subsection{Consistency for the Criterion}
\begin{theorem}
For $\sqrt{n}$-consistent $\widetilde{\bolPhi}$, $\widehat{\mbU}$, and $\widehat{\mbV}$, $\text{Pr}\{(\widehat{s}_{1},\widehat{s}_{2})_{\text{IC}}=(s_{1},s_{2})\} \rightarrow 1$ as $N\rightarrow \infty$.
\label{S-thm:consistencyIC}
\end{theorem}
\noindent \textit{Proof.} To facilitate the proof of Theorem \ref{S-thm:consistencyIC}, we introduce $\sqrt{n}$-consistency for $\bolPhi$ and the subspaces $\mbU$ and $\mbV$ as follows.
\begin{proposition}
By the Central Limit Theorem, $\widetilde{\bolPhi}$ is a $\sqrt{n}$-consistent estimator for $\bolPhi$. Therefore, the eigenvectors and eigenvalues of $\widetilde{\bolPhi}$ are $\sqrt{n}$-consistent for the eigenvectors and eigenvalues of their population counterparts.	
\end{proposition}

We assume the $\sqrt{n}$-consistency for $\widetilde{\bolPhi}$, $\widehat{\mbU}$, $\widehat{\mbV}$. Let $J_{N}(\widehat{\mbU},\widehat{\mbV},s_{1},s_{2})=-\log \big\{\|\mbP_{\widehat{\mbU}}\widetilde{\bolPhi}\mbP_{\widehat{\mbV}}\|_{\mathrm{F}}\big\}$. In the application, we use grid search to determine the $(\widehat{s}_{1},\widehat{s}_{2})$. 
To guarantee our information criteria, we need to show that $$\text{Pr}\Big\{\text{IC}_{N}(\widehat{\mbU},\widehat{\mbV},\widehat{s}_{1},\widehat{s}_{2})-\text{IC}_{N}(\widehat{\mbU},\widehat{\mbV},s_{1},s_{2})>0\Big\} \rightarrow 1, \ \text{as}  \ N \rightarrow \infty$$ 
for the following cases.
$$\begin{cases}
    &\text{(a)} \ 0 < s_{1} \leq \widehat{s}_{1}, \ 0 < s_{2} \leq \widehat{s}_{2} \\
    &\text{(b)} \ 0 < \widehat{s}_{1} < s_{1}, \ 0 < s_{2} \leq \widehat{s}_{2}\\
    &\text{(c)} \ 0 < s_{1} \leq \widehat{s}_{1}, \ 0 < \widehat{s}_{2} < s_{2} \\
    &\text{(d)} \ 0 < \widehat{s}_{1} < s_{1}, \ 0 < \widehat{s}_{2} < s_{2} 
\end{cases}$$
By definition of $\text{IC}_{N}\{s_{1},s_{2}\}$, we have 
\begin{align}
&\text{IC}_{N}(\widehat{\mbU},\widehat{\mbV},\widehat{s}_{1},\widehat{s}_{2})-\text{IC}_{N}(\widehat{\mbU},\widehat{\mbV},s_{1},s_{2}) \nonumber \\
& \hspace{5em} =J_{N}(\widehat{\mbU},\widehat{\mbV},\widehat{s}_{1},\widehat{s}_{2})-J_{N}(\widehat{\mbU},\widehat{\mbV},s_{1},s_{2})+\big\{(\widehat{s}_{1}-s_{1})+(\widehat{s}_{2}-s_{2})\big\}\frac{\log(N)}{N}
\label{S-eq:subs_IC}
\end{align}
First, in (a), by $\sqrt{n}$-consistency, 
$$J_{N}(\widehat{\mbU},\widehat{\mbV},\widehat{s}_{1},\widehat{s}_{2})-J_{N}(\widehat{\mbU},\widehat{\mbV},s_{1},s_{2})=J(\mbU,\mbV,\widehat{s}_{1},\widehat{s}_{2})-J(\mbU,\mbV,s_{1},s_{2})+O_{p}(N^{-1/2}).$$ 
For some $\widehat{s}_{1}$ and $\widehat{s}_{2}$, $J(\mbU,\mbV,\widehat{s}_{1},\widehat{s}_{2})$ converges in probability $J(\mbU,\mbV,s_{1},s_{2})$. Then, 
$$J_{N}(\widehat{\mbU},\widehat{\mbV},\widehat{s}_{1},\widehat{s}_{2})-J_{N}(\widehat{\mbU},\widehat{\mbV},s_{1},s_{2})=J(\mbU,\mbV,\widehat{s}_{1},\widehat{s}_{2})-J(\mbU,\mbV,s_{1},s_{2})+O_{p}(N^{-1/2})=O_{p}(N^{-1/2}).$$
And, it follows from \eqref{S-eq:subs_IC} that the dominant term in $\text{IC}_{N}(\widehat{\mbU},\widehat{\mbV},\widehat{s}_{1},\widehat{s}_{2})-\text{IC}_{N}(\widehat{\mbU},\widehat{\mbV},s_{1},s_{2})$ is 
$$\big\{(\widehat{s}_{1}-s_{1})+(\widehat{s}_{2}-s_{2})\big\}\cdot \frac{\log(N)}{N}, \ \text{which is a positive number.}$$
Next, in (d), since $J_{N}(\widehat{\mbU},\widehat{\mbV},s_{1},s_{2}) < J_{N}(\widehat{\mbU},\widehat{\mbV},\widehat{s}_{1},\widehat{s}_{2})$, it is suffice to show that 
$$J_{N}(\widehat{\mbU},\widehat{\mbV},\widehat{s}_{1},\widehat{s}_{2})-J_{N}(\widehat{\mbU},\widehat{\mbV},s_{1},s_{2})=J(\mbU,\mbV,\widehat{s}_{1},\widehat{s}_{2})-J(\mbU,\mbV,s_{1},s_{2})+o_{p}(1)$$ 
where $J(\mbU,\mbV,s_{1},s_{2})<J(\mbU,\mbV,\widehat{s}_{1},\widehat{s}_{2})<0$. We can decompose 
$$J_{N}(\widehat{\mbU},\widehat{\mbV},i,j)=J(\mbU,\mbV,i,j)+o_{p}(1) \ \text{for all} \ i=1,\ldots,p_{1},j=1,\ldots,p_{2}$$ 
since we have $\sqrt{n}$-consistent $\widetilde{\bolPhi}$, $\widehat{\mbU}$, $\widehat{\mbV}$ and $\mbU$,$\mbV$ affect the $J_{N}(\mbU,\mbV)$ and $J(\mbU,\mbV)$ only through $\mathrm{span}(\mbU)$ and $\mathrm{span}(\mbV)$. In addition, it is straightforward that the sum of two terms that both converge to zero at the same rate converges to zero at the same rate ($o_{p}(1)+o_{p}(1)=o_{p}(1)$).

In (b), it is the same as in (d) when $s_{1} > \widehat{s}_{2}$. Also, the steps are the same as in (a) when $s_{1}<\widehat{s}_{2}$. Similarly, in (c), it is the same as in (a) when $\widehat{s}_{1} > s_{2}$. And, the procedures are the same as in (d) when $\widehat{s}_{1} < s_{2}$.

Therefore, $\text{Pr}\Big\{\text{IC}_{N}(\widehat{\mbU},\widehat{\mbV},\widehat{s}_{1},\widehat{s}_{2})-\text{IC}_{N}(\widehat{\mbU},\widehat{\mbV},s_{1},s_{2})>0\Big\} \rightarrow 1$ as $N \rightarrow \infty$.

\subsection{Simulation on the Criterion}
\begin{figure}[H]
\centering
  \includegraphics[width=0.9\linewidth]{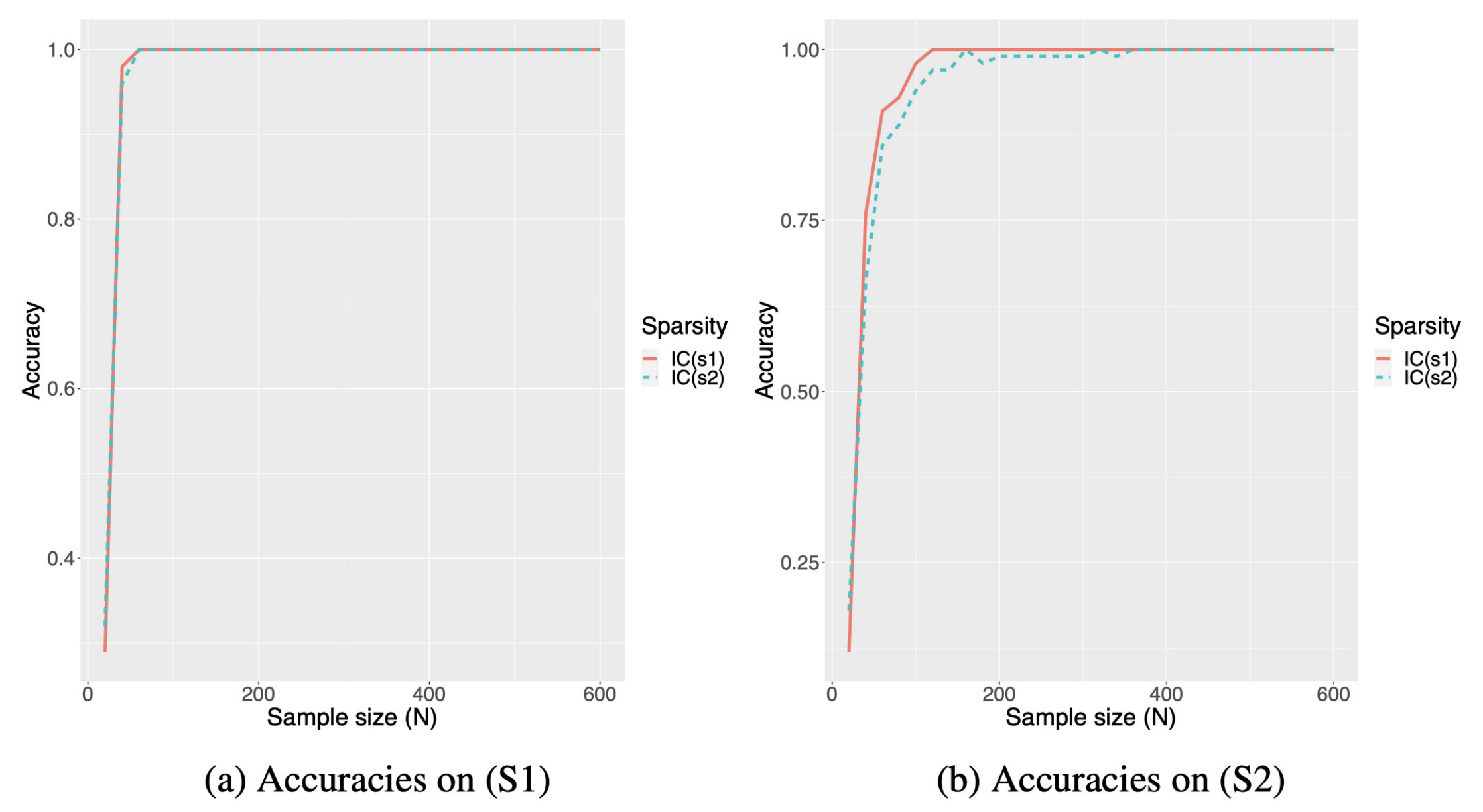}  
\caption{Accuracies of the information criterion (IC) on each scenario with 100 replicates when fixing the true sparsity $s_{1}=s_{2}=3$ and sample size $N \in \{20,40,\ldots, 600\}$.}
\label{S-fig:accuracy}
\end{figure}

In this section, we perform simulations of the information criterion to select sparsity levels $s_{1}$ and $s_{2}$ for the CCR model. First, we fix $p_1=15$, $p_2=15$ and set the true sparsity levels $s_{1}=s_{2}=3$. We apply the same covariance structure in the simulation setup in Section 3.1. And we change the sample size $N=n_{1}+n_{2} \in \{20,40,60,\ldots,600\}$. Thus, we have two multivariate variables $\mbX \in \mathbb{R}^{15}$, $\mbY \in \mathbb{R}^{15}$ and each group has $n_{1}$ and $n_{2}$ observations and we estimate sparsity levels using the information criteria with 100 replicates. In Figure \ref{S-fig:accuracy}, we label each sparsity in different colors. For example, IC(s1) denotes accuracy of $s_{1}$ in the information criterion. The criterion attains an accuracy of 1 after the sample size is larger than 60 (30 in each subgroup). 

\begin{figure}[H]
\centering
  \includegraphics[width=0.45\linewidth]{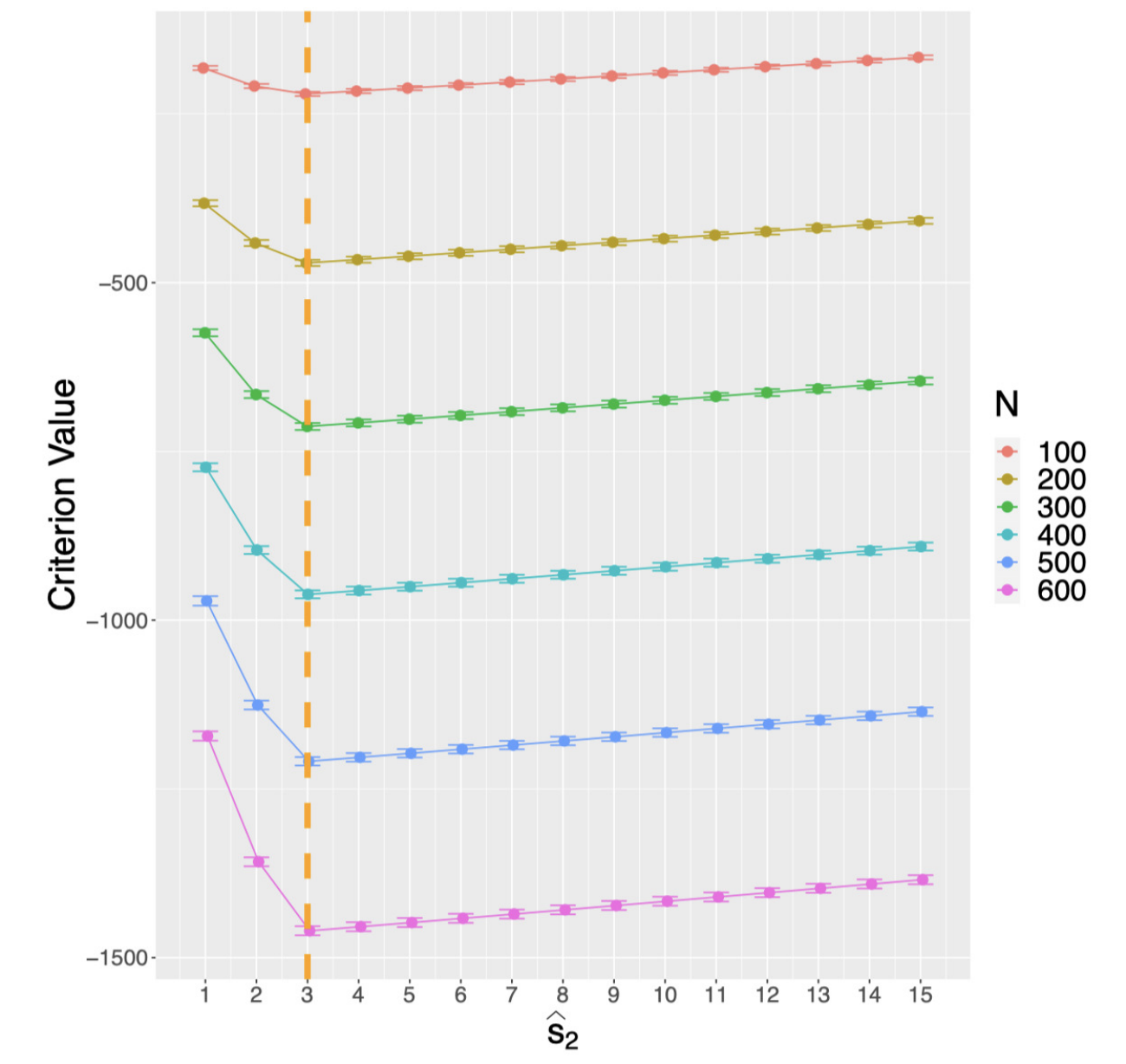}  
\caption{Accuracies of the information criterion (IC) with 100 replicates when fixing the true sparsity $s_{1}=s_{2}=3$ and sample size $N \in \{20,40,\ldots, 600\}$.}
\label{S-fig:criterion_value}
\end{figure}

We explore the behavior of the information criterion with different sample sizes in Figure \ref{S-fig:criterion_value}. We fix $\widehat{s}_{1}=s_{1}=3$ and vary $\widehat{s}_{2}$ from 1 to 15 in each sample size $N$ from 100 to 600. The information criterion achieves the lowest criterion value on $\widehat{s}_{2}=3$ (which is the true $s_{2}$) in all sample sizes. This also supports the criterion is reasonable to determine the number of selected variables in applications when the sample size is large enough ($N \geq 60$).

\section{Resampling Methods for the Sparsity Levels}\label{S-sec:resampling}

We consider two different resampling methods to examine variable selection accuracy under the sparsity assumption. The first is bootstrap sampling. The second is the LTO method, in which one observation from $\mbX$ and one from $\mbY$ are left out in each round, and the CCR model is applied to the remaining $N-2$ observations.

We simulate data similar to rank-1 scenario in Section 3.1. We set $p_{1}=15$, $p_{2}=18$, $s_{1}=3$, $s_{2}=3$, $n_{1}=20$, and $n_{2}=20$. The ratios in Table \ref{S-table:resampling} are the ratios of each variable selected by each resampling method over 1000 replicates with $(s_{1},s_{2})=(3,3)$, that is, the CCR model selects three variables in $\mbX$ and $\mbY$ in each round. The reference ratio is the ratio of variable selection when we randomly select three variables in each of $\mbX$ and $\mbY$. The first three variables in $\mbX$ and $\mbY$ have ratios close to 1, since the selected variables have meaningful signals in $\bolSigma_{\mbX,1}$ and $\bolSigma_{\mbY,1}$. The result demonstrates that the CCR model successfully denoises the nonsignificant signals.

\begin{table}[H]
\centering
\caption{Ratios of selected variables on the CCR model with $\mbX$ and $\mbY$ on the strength of signal is $c_1/c_{2}=3$ where $n_{1}=n_{2}=20$. The reference ratios represent ratios when we randomly select variables on sparsity levels $s_{1}$, $s_{2}$.}
\begin{tabular}{cccccccccccc}
\multicolumn{1}{c}{\multirow{2}{*}{$\mbX$}} & \multicolumn{4}{c}{Bootstrap}                                & \multirow{2}{*}{LTO} & \multicolumn{1}{c}{\multirow{2}{*}{$\mbY$}} & \multicolumn{4}{c}{Bootstrap}                                & \multirow{2}{*}{LTO} \\
\multicolumn{1}{c}{} & 10 & 50 & 100 & 200 & & 
\multicolumn{1}{c}{}  & 10 & 50 & 100 & 200 &  \\ \hline \hline
\multicolumn{1}{c}{1}
& \textbf{0.98} & \textbf{1.00} & \textbf{1.00} & \textbf{1.00} & \textbf{1.00} & 
\multicolumn{1}{c}{1} & \textbf{0.98} & \textbf{1.00} & \textbf{1.00} & \textbf{1.00} & \textbf{1.00} \\
\multicolumn{1}{c}{2} 
& \textbf{0.99} & \textbf{1.00} & \textbf{1.00} & \textbf{1.00} & \textbf{1.00} & 
\multicolumn{1}{c}{2}
& \textbf{0.88} & \textbf{1.00} & \textbf{1.00} & \textbf{1.00} & \textbf{1.00} \\
\multicolumn{1}{c}{3}
& \textbf{0.98} & \textbf{1.00} & \textbf{1.00} & \textbf{1.00} & \textbf{1.00} & 
\multicolumn{1}{c}{3}
& \textbf{0.99} & \textbf{1.00} & \textbf{1.00} & \textbf{1.00} & \textbf{1.00}\\
\multicolumn{1}{c}{4}
& 0 & 0 & 0 & 0 & 0 & 
\multicolumn{1}{c}{4}
 & 0.01 & 0 & 0 & 0 & 0 \\
\multicolumn{1}{c}{5} 
& 0 & 0 & 0 & 0 & 0 & 
\multicolumn{1}{c}{5}                  
& 0.02 & 0 & 0 & 0 & 0\\
\multicolumn{1}{c}{6} 
& 0 & 0 & 0 & 0 & 0 & 
\multicolumn{1}{c}{6}
 & 0.01 & 0 & 0 & 0 & 0\\
\multicolumn{1}{c}{7}                  
& 0 & 0 & 0 & 0 & 0 & 
\multicolumn{1}{c}{7}                  
& 0.01 & 0 & 0 & 0 & 0\\
\multicolumn{1}{c}{8}                  
& 0 & 0 & 0 & 0 & 0 & 
\multicolumn{1}{c}{8}                  
& 0.01 & 0 & 0 & 0 & 0 \\
\multicolumn{1}{c}{9}                  
& 0 & 0 & 0 & 0 & 0 & 
\multicolumn{1}{c}{9}                  
& 0 & 0 & 0 & 0 & 0 \\
\multicolumn{1}{c}{10}                 
& 0.01 & 0 & 0 & 0 & 0 & 
\multicolumn{1}{c}{10}                 
& 0 & 0 & 0 & 0 & 0\\
\multicolumn{1}{c}{11}                   
& 0 & 0 & 0 & 0 & 0 &  
\multicolumn{1}{c}{11}                 
& 0.04 & 0 & 0 & 0 & 0 \\
\multicolumn{1}{c}{12}                   
& 0 & 0 & 0 & 0 & 0 & 
\multicolumn{1}{c}{12}                 
& 0 & 0 & 0 & 0 & 0 \\
\multicolumn{1}{c}{13}                   
& 0.01 & 0 & 0 & 0 & 0 & 
\multicolumn{1}{c}{13}                 
& 0 & 0 & 0 & 0 & 0 \\
\multicolumn{1}{c}{14}                   
& 0.01 & 0 & 0 & 0 & 0 & 
\multicolumn{1}{c}{14}                 
& 0 & 0 & 0 & 0 & 0 \\
\multicolumn{1}{c}{15}                   
& 0.02 & 0 & 0 & 0 & 0 & 
\multicolumn{1}{c}{15}                 
& 0.01 & 0 & 0 & 0 & 0 \\
\multicolumn{1}{c}{}                   
& & & & & & 
\multicolumn{1}{c}{16}                 
& 0.01 & 0 & 0 & 0 & 0 \\
\multicolumn{1}{c}{}                   
& & & & & & 
\multicolumn{1}{c}{17}                 
& 0.02 & 0 & 0 & 0 & 0 \\
\multicolumn{1}{c}{}                   
& & & & & & 
\multicolumn{1}{c}{18}                 
& 0.01 & 0 & 0 & 0 & 0 \\ \hline
\multicolumn{5}{c}{Reference ratio}               & 0.20 & 
\multicolumn{5}{c}{Reference ratio}                & 0.17   \end{tabular}
\label{S-table:resampling}  
\end{table}

\section{Computational Complexity}\label{S-sec:computational_complexity}
For the computational complexity, let $\mathbf{X} \in \mathbb{R}^{N \times p_{1}}$ and $\mathbf{Y} \in \mathbb{R}^{N \times p_{2}}$ with total sample size $N=n_{1}+n_{2}$. First, the computational complexity for calculating the cross-covariance is  $\mathbf{\Sigma}_{\mathbf{XY}}(z) \in \mathbb{R}^{p_{1} \times p_{2}}$, the computational complexity is $\mathcal{O}(Np_{1}p_{2})$. Then, we need to compute the singular value decomposition (SVD) of $\mathbf{\Phi}=\mathbf{\Sigma}_{\mathbf{XY}}(1)-\mathbf{\Sigma}_{\mathbf{XY}}(2) \in \mathbb{R}^{p_{1} \times p_{2}}$ and  the computational complexity is $\mathcal{O}(\min(p_{1}^{2}p_{2},p_{1}p_{2}^{2}))$. Lastly, for a sparse SVD for variable selection, the computational complexity is $\mathcal{O}(p_{1}p_{2}r)$ when $r$ is a reduced rank. Thus, the total complexity of the CCR model is $\mathcal{O}(Np_{1}p_{2})+\mathcal{O}(\min(p_{1}^{2}p_{2},p_{1}p_{2}^{2}))+\mathcal{O}(p_{1}p_{2}r)$.
For small $p_{1}$ and $p_{2}$, this simplifies to $\mathcal{O}(Np_{1}p_{2})$, as matrix multiplications and SVD dominate. When $p_{1}$ and $p_{2}$ are large ($\gg N$), these complexities approximate $\mathcal{O}(p_{1}^{3})$ when $p_{1} \approx p_{2}$.

\section{Real Data Analysis}\label{S-sec:further_analysis}
\subsection{Variables for Analysis of Skull and Temporalis Origin Muscle}
    
\begin{table}[H]
\centering
\caption{Variables in skull and temporalis origin muscle (TO) with 10 male and 11 female subjects. The skull has 21 subjects with 16 variables and the temporalis origin (TO) has 21 subjects with 18 variables. The asterisks represent the bilaterally measured variables.}
\begin{tabular}{cccl}
& Attribute & Variable & Descriptions and measurement \\ \hline \hline 
\multirow{16}{*}{Skull}                                                                     & \multirow{9}{*}{\begin{tabular}[c]{@{}c@{}}Feature \\ points\end{tabular}}     
& GnToGn & distance between left \& right gonions\\
& & \cellcolor{Gray} AnL & \cellcolor{Gray}distance between most anterior \& posterior poles\\
& & AnH & distance between most superior \& inferior poles\\
& & \cellcolor{Gray} AnT & \cellcolor{Gray}distance between most medial \& lateral poles\\
& & PlToPr & 
\begin{tabular}[c]{@{}l@{}}distance between the most lateral points of \\ the roofs of the left(Pl) \& right(Pr) ear canals\end{tabular}  \\
& & \cellcolor{Gray} NaToPlPr & \cellcolor{Gray}\begin{tabular}[c]{@{}l@{}}vertical distance from nasal bone suture to the \\ line connecting left and right ear canals(Pl \& Pr)\end{tabular} \\
& & $\text{CrToGn}^{*}$ & distance from coronoid process to gonion \\ \cline{2-4}
& \multirow{2}{*}{Ramus}
& \cellcolor{Gray}$\text{Length}^{*}$ & \cellcolor{Gray}\begin{tabular}[c]{@{}l@{}}distance from the highest point on the \\ mandibular condyle to the gonion\end{tabular} \\
& & $\text{Width}^{*}$ & least width perpendicular to the ramus length\\ \cline{2-4}
& \multirow{4}{*}{Mandible}
& \cellcolor{Gray}$\text{Length}^{*}$ & \cellcolor{Gray}\begin{tabular}[c]{@{}l@{}}distance from the highest point on the   \\mandibular condyle to the anterior margin of chin\end{tabular} \\
& & $\text{Angle}^{*}$ & \begin{tabular}[c]{@{}l@{}}angle formed by tangent between the lower border\\ of the mandible and the posterior border of the\\  ramus from the condyle to gonion \end{tabular}                             \\ \hline
\multirow{10}{*}{TO} 
& \multirow{5}{*}{Size}
& \cellcolor{Gray} $\text{BoxLength}^{*}$ & \cellcolor{Gray}length of 3D box on the symmetric plane\\
& & $\text{BoxWidth}^{*}$ & width of 3D box on the symmetric plane \\
& & \cellcolor{Gray} $\text{BoxThickness}^{*}$ & \cellcolor{Gray}thickness of 3D box parallel to symmetric plane                              \\
& & $\text{Area}^{*}$ & actual muscle attachment surface \\
& & \cellcolor{Gray} $\text{Volume}^{*}$ & \cellcolor{Gray}actual volume measured by determination kit                              \\ \cline{2-4}
& \multirow{3}{*}{\begin{tabular}[c]{@{}c@{}}Spatial\\ orientation\end{tabular}} 
& $\text{SA}^{*}$ & angle between box plane and sagittal plane \\
& & \cellcolor{Gray} $\text{FA}^{*}$ & \cellcolor{Gray}angle between box plane and frontal plane \\
& & $\text{FHA}^{*}$ & angle between box plane and Frankfurt plane                              \\ \cline{2-4}
& Location
& \cellcolor{Gray} \begin{tabular}[c]{@{}c@{}}Centroid\\ $\text{Distance}^{*}$\end{tabular} & \cellcolor{Gray}distance from origin to centroid of the muscle                             
\end{tabular}
\label{S-table:variables}
\end{table}

\subsection{Real Data Analysis on Skull and Temporalis Origin}\label{S-sec:skullvsTO}

For additional biomechanical interpretation, we further display mandibular length (MandibleLength (R)) and attachment area (Area(L)) in (5) of the manuscript, which highlight differences in marginal covariance, as shown in Figure 5 of the main manuscript.

\begin{figure}[H]
\centering
  \includegraphics[width=0.65\linewidth]{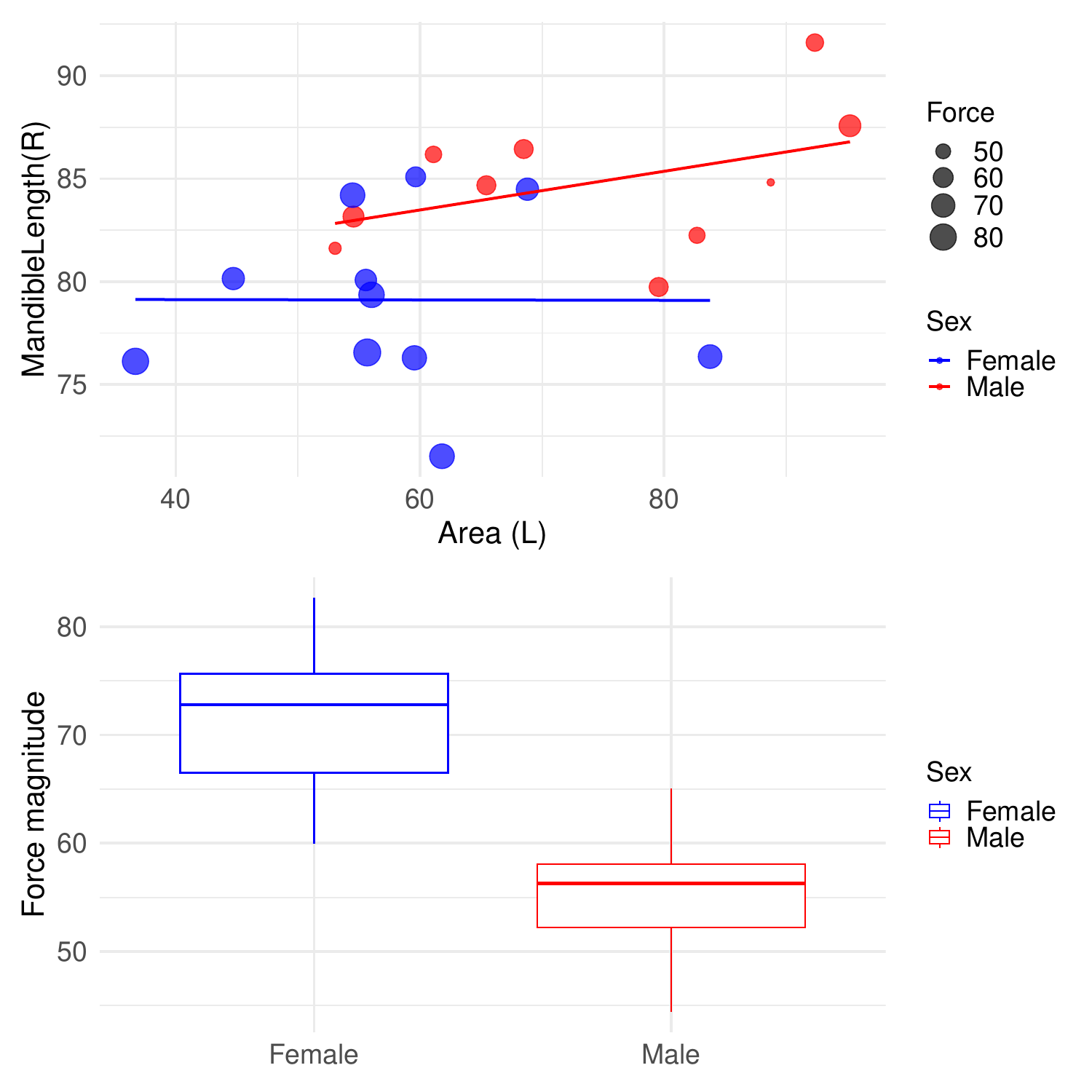}  
\caption{Plots of mandibular length (MandibleLength(R)), and the left temporalis origin surface areas (Area (L)), and joint reaction force magnitude in raw-scale measurements.}
\label{S-fig:plot_skull_vs_TO_areaL}
\end{figure}

Figure \ref{S-fig:plot_skull_vs_TO_areaL} shows a scatter plot of mandibular length (MandibleLength (R)) and temporalis origin (TO) attachment surface areas (Area (L)) with joint reaction force (JRF) magnitude on the raw scale (top), and a boxplot of JRF magnitude by sex (bottom). The JRF vector is calculated as the residual force at the TMJ required to maintain static equilibrium under estimated muscle forces during mandibular motion \citep{she2021sexual}, and its magnitude is defined as the length of this vector. A larger JRF corresponds to greater loading on the TMJ. For the same level of bite force, JRF increases as 3D mandibular length decreases \citep{sun2024explainable}.

In the raw-scale plot shown in Figure \ref{S-fig:plot_skull_vs_TO_areaL}, females with shorter mandibular length (MandibleLength (R)) and smaller attachment areas (Area (L)) exhibit larger force magnitudes (top), and the joint reaction force in females is substantially greater than in males (bottom). Notably, our analysis using the CCR model indicates that females tend to have a shorter mandibular length (MandibleLength (R)) as the selected size-related variables in TO increase, while other selected skull variables are held fixed. This finding may help identify a high-risk subgroup of females for TMD, and the association appears faint in the raw-scale plot, which only reflects pairwise variable relationships, as shown in Figure \ref{S-fig:plot_skull_vs_TO_areaL}.

Thus, the CCR model provides new insights into the association between the linear combinations of the skull and muscle attachment. This suggests a possible future research direction focusing on abnormal or high-risk samples to further investigate the association between bone structure and muscle measurements.

\subsection{Real Data analysis on Marginally Standardized Skull and Temporalis Origin}\label{S-sec:skullvsTO_standardized}
This section presents CCR model results with marginally standardized skull and temporalis origin (TO) muscle measurements. Marginal standardization refers to standardizing within each group of $Z$. We use $(s_{1},s_{2})=(3,3)$ based on the results of the SPSS method. The standardized vectors and datasets are denoted with the subscript $s$.

\begin{figure}[H]
\centering
  \includegraphics[width=0.5\linewidth]{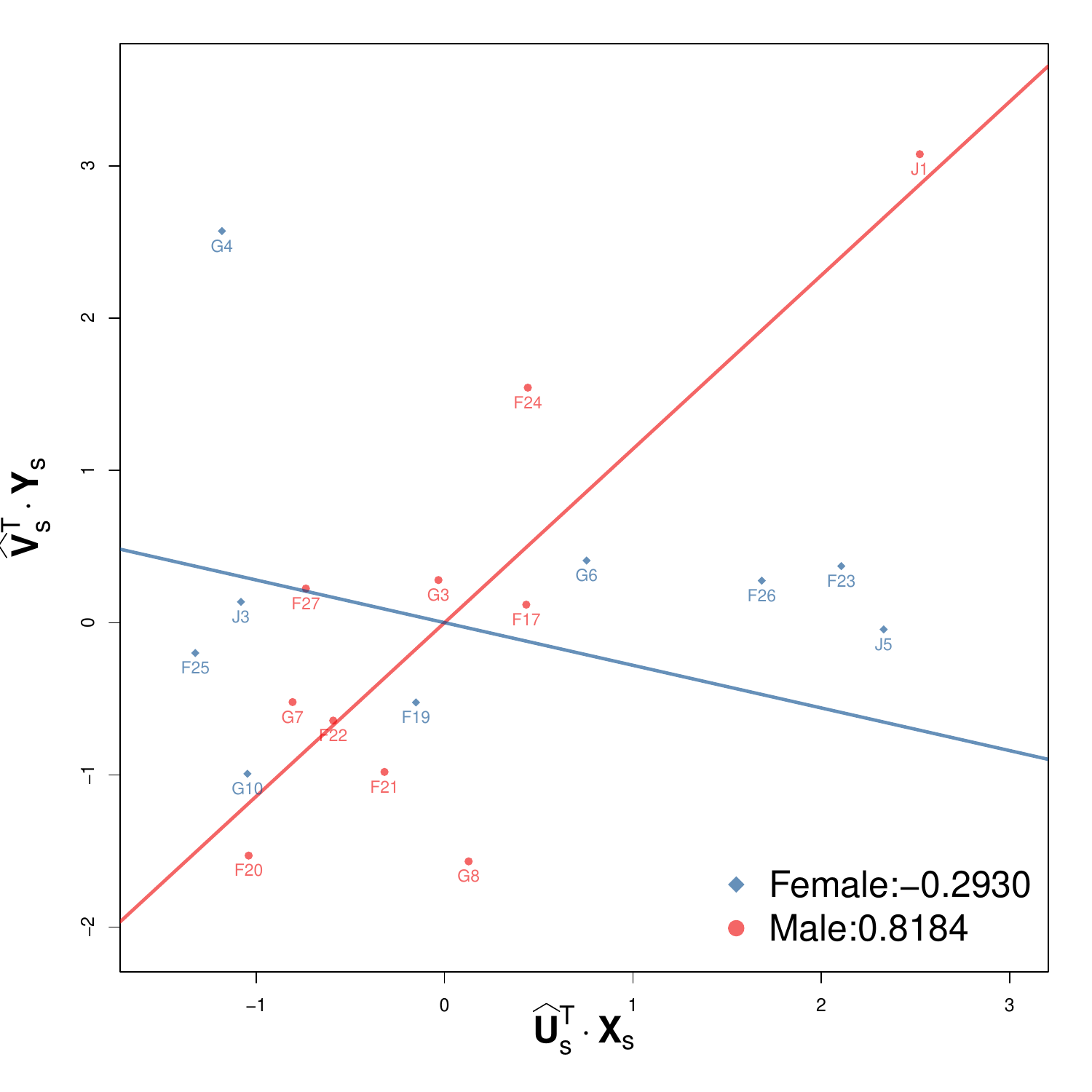}  
\caption{Linear combinations of the CCR model on the marginally standardized skull and temporalis origin (TO) measurements. The numbers in the legend represent the correlations between linear combinations by sex. Sparsity levels are set as $s_{1} = s_{2} = 3$. The $x$-axis ($\widehat{\mbU}_{s}^{\top}\mbX_{s}$) indicates the linear combination on the standardized skull. The $y$-axis ($\widehat{\mbV}_{s}^{\top}\mbY_{s}$) represents the linear combination on the standardized temporalis origin (TO).}
\label{S-fig:skull_vs_TO_std_plot}
\end{figure}

\begin{figure}[H]
\centering
  \includegraphics[width=0.8\linewidth]{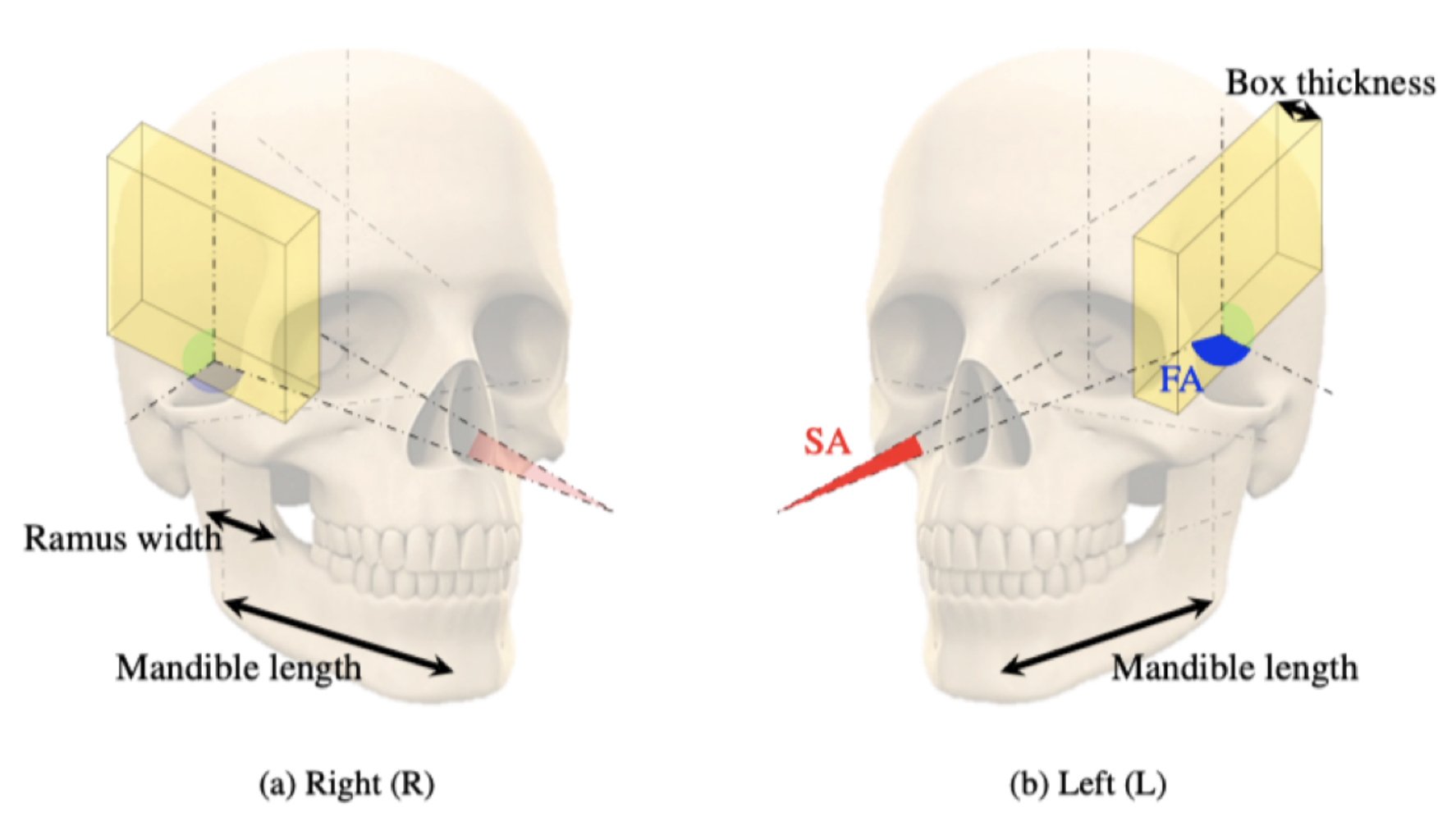}  
\caption{Result of the CCR model on the marginally standardized skull and temporalis origin (TO) under $(s_{1},s_{2})=(3,3)$ where selected 6 variables are denoted with variable names.}
\label{S-fig:skull_vs_TO_std_skullfig}
\end{figure}

The estimated maximal covariance difference and associated correlation differences are $\widehat{\delta}_{s}=1.96$ and  $\widehat{\eta}_{s}=1.11$, respectively. In \eqref{S-eq:selected_variables_standardized_skullvsTO}, we can see that left and right mandible lengths are selected on the standardized skull, and the angle variables (SA, FA) are selected on the standardized TO. The result is different from the non-standardized results in Section 4 since the marginal standardization removed the scale effects. However, since we could not obtain biomechanical evidence supporting these linear combinations in \eqref{S-eq:selected_variables_standardized_skullvsTO}, only the non-standardized results are included in Section 4. 
\begin{align}
\text{Skull:}\enspace \widehat{\mbU}_{s}^{\top} \mbX_{s}  &= 0.443 \enspace \text{Mandible.Length(L)} +0.524  \enspace \text{Ramus.Width(R)} \nonumber \\ &+0.726 \enspace \text{Mandible.Length(R)}\nonumber \\
\text{TO:} \enspace \widehat{\mbV}_{s}^{\top}\mbY_{s}  &= 0.387 \enspace \text{BoxThickness(L)} -0.602 \enspace \text{SA(L)} +0.697 \enspace \text{FA(L)}
\label{S-eq:selected_variables_standardized_skullvsTO}
\end{align}
Figure \ref{S-fig:skull_vs_TO_std_plot} displays the result of the CCR model in plots of the linear combinations, and the corresponding graphical example of the standardized skull and TO muscle are displayed in Figure \ref{S-fig:skull_vs_TO_std_skullfig}.

\label{lastpage}

\end{document}